\documentclass[twocolumn]{aastex63}
\usepackage{multirow}
\usepackage{booktabs}
\usepackage{amsmath}
\usepackage{threeparttable}
\usepackage{hyperref}			
\newcommand{\customcite}[2]{\hyperlink{cite.#1}{#2}}
\hypersetup{colorlinks,
	linkcolor=blue,
	anchorcolor=blue,
	citecolor=blue}
\newcommand{\sersic}{S\'{e}rsic}

\newcommand{\RNum}[1]{\uppercase\expandafter{\romannumeral #1\relax}}
\usepackage{graphicx}
\shorttitle{Morphology of galaxies in TNG100 simulations}
\shortauthors{Gong et al.}
\graphicspath{{./}{figures/}}
\submitjournal{ApJS}
\begin{document}
	
	\title{Mock Observations: Morphological Analysis of Galaxies in TNG100 Simulations}
	
	\correspondingauthor{Lin, Weipeng}
	\email{linweip5@mail.sysu.edu.cn}
	
	\author{Jun-Yu Gong}
	\affiliation{School of Physics and Astronomy, Sun Yat-sen University, DaXue Road 2, 519082, Zhuhai, China}
	\affiliation{CSST Science Center for the Guangdong-Hongkong-Macau Greater Bay Area, DaXue Road 2, 519082, Zhuhai, China}
	
	\author{Weipeng Lin}
	\affiliation{School of Physics and Astronomy, Sun Yat-sen University, DaXue Road 2, 519082, Zhuhai, China}
	\affiliation{CSST Science Center for the Guangdong-Hongkong-Macau Greater Bay Area, DaXue Road 2, 519082, Zhuhai, China}
	
	\author{Lin Tang}
	\affiliation{CSST Science Center for the Guangdong-Hongkong-Macau Greater Bay Area, DaXue Road 2, 519082, Zhuhai, China}
	\affiliation{School of Physics and Astronomy, China West Normal University, ShiDa Road 1, 637002, Nanchong, China}
	
	\author{Yanyao Lan}
	\affiliation{School of Physics and Astronomy, Sun Yat-sen University, DaXue Road 2, 519082, Zhuhai, China}
	\affiliation{CSST Science Center for the Guangdong-Hongkong-Macau Greater Bay Area, DaXue Road 2, 519082, Zhuhai, China}
	\affiliation{School of Optoelectronic Engineering, Guangdong Polytechnic Normal University, Guangzhou 510665, PR China}
	
	\begin{center}
		\begin{abstract}
			
			In this study, we investigate the morphology of galaxies in the TNG100 simulation by applying mock observation techniques and compare the results with the observational data from the Sloan Digital Sky Survey (SDSS). By employing a hierarchical Convolutional Neural Network (CNN) approach, we classify galaxies into four morphological categories (Ellipticals, S0/a, Sab/Sb, and Sc/Sd/Irregulars). Our findings show that the morphological characteristics of the mock-observed galaxy samples closely match those observed in the SDSS, successfully reproducing key features such as distinct parameter distributions for different types. However, some discrepancies are identified: notably, a significant lack of early-type galaxies (ETGs) in the dwarf galaxy regime ($M_* < 10^{10} M_{\odot}$) and minimal distinction between Sab/Sb and Sc/Sd/Irregular galaxies in the mock-observed samples, unlike the clear differences seen in actual observations. These divergences may stem from simulation properties such as elevated star formation efficiency at low mass end or resolution limits. Observational effects, including the impact of the Point Spread Function, sky background, and instrumental noise, can independently cause approximately 7.87\% morphological misclassifications by our CNN model. Compared to previous studies using gravity-based definitions of galaxies that failed to clearly distinguish the parameter distributions of ETGs versus Late-Type Galaxies, our brightness-based sample definition method better recovers the observed morphological parameter distributions, especially their distinct differences. Our study underscores that, alongside mock observations, employing galaxy segmentation methods consistent with observational practices is crucial for appropriately recovering realistic morphological parameters from simulations and enabling fair comparisons. A catalog with morphological parameters is provided for further analysis.
			
		\end{abstract}
	\end{center}
	
	\keywords{galaxies: formation, galaxies: photometry, galaxies: statistics, galaxies: structure}
	
	\section{Introduction}\label{Sec_Intro}
	The morphology of galaxies is a fundamental aspect significantly influenced by their formation processes and evolutionary history. Galaxies are commonly categorized into elliptical (early-type) and disk (late-type) categories. Elliptical galaxies, typically formed through short-term, violent interactions such as major mergers (\citealt{Toomre1972, Barnes1992, Ceverino2010}), exhibit smooth structures with stellar motions predominantly characterized by high velocity dispersion. Their stellar populations are generally older and exhibit redder colors (see \citealt{Bell2004, Thomas2005, DeLucia2006, Lintott2008, Huang2016, Lacerna2020}). In contrast, disk galaxies, which evolve through secular processes and minor mergers (\citealt{Dekel2003, Kormendy2004, Sheth2005}), demonstrate stellar systems dominated by rotational motion. These galaxies often display distinct substructures such as spiral arms, rings, and breaks. The stellar populations in disk galaxies are comparatively younger and bluer (see \citealt{Brinchmann2004, Kormendy2004, Schreiber2009, Kormendy2013, Leroy2013}). This diversity in galaxy morphology underscores the various pathways of formation and evolution, pointing out the importance of morphological studies in comprehending the stellar composition and evolutionary trajectories of galaxies (\citealt{Fasano2000, Volonteri2007, van2008, Tacchella2019}).
	
	Numerical simulations are an indispensable tool in the study of galaxy morphology. They provide a unique perspective, enabling the exploration of a myriad of cosmic processes, ranging from the formation of large-scale structures in the universe to the detailed internal dynamics of galaxies on smaller scales (\citealt{Furlong2015, Snyder2015, Haider2016, Lagos2017}). The TNG project, building upon the foundation of the original Illustris project, represents a significant advancement (\citealt{Nelson_2018, Naiman_2018, Du2019, Pulsoni2020, Du2021, Piotrowska_2022, Zana2022}). With its enhanced resolution, sophisticated physical models, and increased simulation volume, TNG provides a more accurate depiction of the cosmic evolutionary processes (\citealt{Lovell2018,Diemer2019,Piotrowska2022}). Notably, TNG has successfully simulated complex galaxy properties, including morphology, color, and star formation history (\citealt{Huertas-Company_2019, Zanisi_2021, Varma_2022}). Furthermore, it features the crucial role of feedback from active galactic nuclei (AGN) in heating or expelling the surrounding gas, thereby shutting down star formation in massive galaxies (\citealt{Quai_2021, Donnari_2021, Pillepich_2018}). These advancements have significantly improved our understanding of the evolutionary paths of galaxies.
	
	Recent studies utilizing the TNG simulations have provided significant insights into the morphological diversity of galaxies and their correlation with various galaxy properties. \cite{Nelson_2018} indicated the morphological diversity across a range of stellar masses and structural characteristics within TNG galaxies, demonstrating the simulation's capability to replicate the variety observed in the real universe. Building on this,  \cite{Rodriguez-Gomez_2019} and \cite{Huertas-Company_2019} further refined the understanding of galaxy morphologies by generating and analyzing synthetic images of galaxies, which were compared with observations from the Pan-STARRS and Sloan Digital Sky Survey (SDSS), respectively. They noted good agreement in optical morphologies and non-parametric morphological diagnostics between TNG galaxies and observations, although some challenges remain, such as a higher fraction of red disk galaxies and blue spheroids in TNG compared to observations, and the fact that the simulation does not reproduce the observed tendency for disk galaxies to be larger than spheroidal systems at fixed stellar mass. Additionally, \cite{Tacchella2019} explored the relationship between galaxy morphology, star formation, and kinematical properties in TNG, revealing correlations between the fraction of stellar mass in a spheroidal component and both stellar mass and star formation activity, in broad agreement with observations. These comprehensive analyses emphasize the importance of systematic comparisons between simulated and observed data, providing crucial insights for refining theoretical models of galaxy formation.
	
	While simulation data provide profound insights into galaxy formation and evolution, the consistency of methodologies is crucial for reliable comparisons between simulations and actual observations. \cite{Pillepich_2018} focused on refining galaxy outflows and black hole feedback mechanisms, successfully aligning the galaxy stellar mass functions with observed data. However, in this study, galaxies were defined to include stellar particles within an arbitrary aperture toward the center of the stellar substructure. This definition clearly differs from that used in observational studies. To facilitate a fair comparison, \cite{Tang_2021} developed a novel approach for producing mock observed galaxies from the TNG100 simulation, utilizing mock imaging and standard observational techniques to extract structural parameters. This approach not only yielded realistic surface-brightness distributions and morphological features but also closely matched the galaxy luminosity and stellar mass functions. 
	
	Recent studies have further leveraged TNG data to produce mock observations, enhancing the alignment between simulations and empirical data. For example, \cite{Byrohl_2021} applied a new radiative transfer code to simulate Lyman alpha haloes in the TNG50 simulation, providing insights into their physical origins and environmental interactions. \cite{Snyder_2023} synthesized galaxy survey fields for the James Webb Space Telescope (JWST) and Hubble Space Telescope, aiding in studies of galaxy evolution. \cite{Sarmiento_2023} generated a mock Mapping Nearby Galaxies at APO (MaNGA) survey from TNG data, accurately mirroring observations of stellar populations and kinematics. Utilizing the same approach in \cite{Tang_2021}, \cite{Lan2024} employed the TNG100 simulation results to predict galaxy alignments, finding that the orientation of the brightest group galaxy strongly correlates with the spatial distribution of its satellites, a signal broadly consistent with observations and revealing the complex interplay between galaxy properties and their distribution within the dark matter halo. These efforts emphasize the essential role of using observation-driven approaches in simulation analyses to generate realistic predictions for comparison with actual observations. 
	
	Building on the aforementioned studies, we aim to investigate the TNG100 simulation data to elucidate the relationships between galaxy properties across different morphologies. Firstly, the TNG100 data will be adjusted to resemble the SDSS observational characteristics by incorporating Point Spread Function (PSF) convolution, adding background noise, and redefining galaxies using the Surface Brightness Limited Segmentation Procedure (SBLSP) method \citep{Tang_2020,Tang_2021}. Using a hierarchical approach with three convolutional neural networks (CNNs), we classify galaxies into four morphological categories: elliptical, S0/a, Sab/Sb, and Scd. Subsequently, we conduct statistical analyses to explore the properties of these morphological classes. The paper will be organized as follows: Section \ref{Sec_data} outlines the data sources, encompassing both numerical simulation data and real observational data. In Section \ref{Sec_method}, we elaborate on our methodology, including data processing techniques and the estimation of various properties for both simulated and observed galaxies. Section \ref{Sec_result} presents the main results, which are subsequently discussed in Section \ref{Sec_disscussion}. Finally, Section \ref{Sec_summary} shows the key findings and conclusions of this study, emphasizing our contributions to the understanding of galaxy morphology from both simulated and observational perspectives.
	
	\section{Data}\label{Sec_data}
	\subsection{Simulations}\label{Sec_data_sim}
	The TNG\footnote{https://www.tng-project.org/} project, a successor to the original Illustris\footnote{https://www.illustris-project.org/} simulations, stands as a monumental endeavor in the realm of hydrodynamic cosmological simulations. This suite of simulations has successfully reproduced many observational properties of galaxies and their fundamental physical relations (\citealt{Marinacci_2018, Naiman_2018, Nelson_2018, Pillepich_2018, Springel_2018}). Notably, the mass-size relationships of both early-type galaxies (ETGs) and late-type galaxies (LTGs) in TNG simulations are remarkably consistent with those in observations (\citealt{Genel_2018, Huertas-Company_2019, Rodriguez-Gomez_2019}).
	
	In this study, we will ultilize the TNG100-1 run, a pivotal component of the TNG suite. This run traces the evolution of $2 \times 1820^3$ resolution elements within a cubic box with a side length of 110.7 Mpc, yielding an average baryonic mass resolution of  $1.39 \times 10^6 M_\odot$. The gravitational softening length designated for stellar entities is set to 0.74 kpc. Galaxy groups and clusters in the simulation are identified using the friend-of-friend (FoF) algorithm (\citealt{Davis1985}).  
	
	\subsection{Observations}\label{SDSS_data}
	The SDSS has been an invaluable resource for astronomical research, providing extensive spectral and photometric data across a significant portion of the sky. The SDSS's dedicated 2.5-meter telescope, located at the Apache Point Observatory, employs a 120-megapixel camera to capture images through five broad-band filters ($u, g, r, i, z$) over 14,555 square degrees of the sky. The survey's innovative design and operation have enabled detailed observations of millions of objects, offering a comprehensive view of the universe's structure and composition (\citealt{York2000,Gunn2006}).
	
	In conjunction with the SDSS, the study by \cite{Nair_2010} (hereafter \customcite{Nair_2010}{N10}) plays a pivotal role in this work by providing a catalog of detailed visual classifications for 14,034 galaxies from the SDSS DR4\footnote{All SDSS data used in this work were downloaded from DR12.}. This catalog  encompasses a substantial portion of galaxies within the redshift range ($0.01 < z < 0.1$) and includes those with an extinction-corrected apparent magnitude limit of ($g < 16$). It documents morphological features such as T-types, bars, rings, lenses, tails, warps, dust lanes, arm flocculence, and multiplicity. Additionally, the catalog records other galaxy properties like magnitudes, stellar mass, star formation rates, AGN types, and group information, making it highly suitable for studies exploring the relationship between galaxy morphology and their formation and evolution. The classification methodology, systematics, biases, and statistical properties of the sample are thoroughly documented, providing a robust foundation for subsequent studies on galaxy morphology and serving as a valuable baseline for the development of automated galaxy classification algorithms (\citealt{Domnguez2018,Huertas-Company_2019}).
	
	\section{Methodology}\label{Sec_method}
	\subsection{TNG100 Galaxy Samples}\label{Sec_tl_data}
	In this study, the definition of simulated galaxies is rooted in the SBLSP algorithm, as elaborated in \cite{Tang_2020,Tang_2021}. We provide a brief summary of the method here, and recommend interested readers refer to the aforementioned articles for a comprehensive explanation. The procedure begins by projecting the stellar particles within the FoF groups onto three planes ($x\text{-}y$, $y\text{-}z$, $z\text{-}x$), where the FoF groups, as mentioned in Section~\ref{Sec_data_sim}, represent galaxy groups or clusters identified using the FoF algorithm. The projected FoF group is divided into a grid of cells, with each cell corresponding to a pixel in a CCD image. For every grid cell, the properties of the stellar particles it contains are calculated: luminosity, stellar mass, and star formation rate (SFR) are obtained by summing the values of all particles, while stellar age and metallicity are determined as mass-weighted averages.
	
	The TNG100 simulation provides detailed stellar population information for each mesh, including luminosities, stellar mass, SFR, age, and metallicity, which are essential for deriving key astrophysical properties such as surface brightness and total stellar mass. The surface brightness in any band $x$ for each pixel is given by
	\begin{equation}\label{equ:mag}
		\mu_x = -2.5 \log \left( \frac{I_x}{L_{\odot,x} \, \text{pc}^{-2}} \right) + 21.572 + M_{\odot,x}, 
	\end{equation}
	where $ L_{\odot,x} $ is the solar luminosity and $ M_{\odot,x} $ is the absolute magnitude of the Sun in the $x$ band. $ I_x$ is defined as
	\begin{equation}\label{equ:flux}
		I_x = \frac{L_x}{\pi^2 D^2 (1 + z)^4}, 
	\end{equation}
	where $L_x$ represents the luminosity in a given grid size  $D$ (set as the softening length) in the $x$ band, and $z$ is the redshift. In the TNG100 simulation, detailed stellar population data for each mesh enable the derivation of key astrophysical quantities across three specific redshifts: $z = 0.01, 0.1$, and $0.2$, within four SDSS bands ($g, r, i, z$). For each image corresponding to a given redshift and band, our processing steps involve: (1) calculating the total stellar mass and luminosity by summing the masses and luminosities of all star particles in the mesh; (2) determining the average age and metallicity of each mesh by mass-weighted calculations; and (3) integrating the star-formation rates of all gas cells within the mesh. After calibrating these images to a pixel scale consistent with the SDSS CCD pixel size of 0.396 arcsec, we take into account the gravitational softening length of the simulation. Notably, at $z = 0$, this softening length of $D = 0.74~\text{kpc}$ is over ten times the SDSS pixel size. However, as we move to higher redshifts, the softening length becomes more comparable to the pixel size. For example, at redshifts $z = 0, 0.1$, and $0.2$, the softening length corresponds to angular scales of $5^{\prime\prime}, 0^{\prime\prime}\!\!.5$, and $0^{\prime\prime}\!\!.3$, respectively.
	
	Defining galaxy samples is essential for comparing theoretical models with observational data. As described in Section \ref{Sec_data_sim}, galaxy systems are identified based on groups or clusters detected using the FoF algorithm. However, within a single FoF group, multiple galaxies may exist, and different methods can be employed to distinguish individual galaxies. In the IllustrisTNG simulation, galaxies are typically identified using the \texttt{subfind} algorithm, which locates self-bound subhalos. Nevertheless, \cite{Tang_2021} demonstrated that defining galaxies based on their surface brightness produces galaxy stellar mass and luminosity functions that align more closely with observations. In observational astronomy, galaxy segmentation commonly assigns brighter pixels to galaxies and fainter pixels to the background; algorithms such as \texttt{SExtractor} \citep{Bertin1996} and \texttt{Photutils} \citep{photutils2020} are often used to perform this task. These tools can also effectively separate close or interacting systems, reducing mutual contamination. Motivated by these findings, we adopt the SBLSP method proposed in \cite{Tang_2021}. A brief overview of our adapted SBLSP steps is provided here, with a more comprehensive discussion available in \cite{Tang_2020, Tang_2021}. In our implementation, we apply a series of surface-brightness limits (SBLs) ranging from 18 to 30 mag arcsec$^{-2}$ in increments of 0.1 mag arcsec$^{-2}$. The upper limit of 30 mag arcsec$^{-2}$ ensures that we fully capture the faint outskirts of galaxies, even though these pixels may ultimately be dominated by noise in subsequent mock observations; retaining them is essential for following realistic mock observation. For each SBL, galaxies are defined as distinct regions of connected pixels that meet or exceed the given brightness threshold. We then compare the galaxy catalogues derived from adjacent SBLs: if a galaxy identified at a fainter SBL splits into multiple systems at a brighter SBL, the fainter-SBL catalogue is updated to reflect these subdivisions. This process is well illustrated in Figure 2 and Figure 3 of \cite{Tang_2020}. By iterating this procedure across all SBLs, we construct galaxy samples for each redshift and projection plane. As noted by \cite{Tang_2021}, the SBLSP method imposes stricter criteria than \texttt{SExtractor}, allowing for more effective separation of closely interacting galaxies.
	
	Dust extinction significantly affects galaxy luminosities, a critical factor in generating realistic mock observations. To account for dust in our mock observations of star-forming galaxies, we apply the radiative transfer code SKIRT \citep{Baes_2011}, adopting the dusty disk model from \cite{Yuan_2021} and the \cite{Calzetti_2000} attenuation curve. For each galaxy, we first compute the axis ratio (b/a) at the 26.5 mag arcsec$^{-2}$ isophotal contour, which represents a surface brightness level near the reliable detection limit in the SDSS $r-$band. This threshold is chosen to balance the inclusion of most galaxy light with minimizing noise contamination, and we use it as a morphological indicator of orientation—face-on (b/a $\approx$ 1) or edge-on (b/a $ \approx$ 0). Based on this orientation, SKIRT calculates the dust-induced attenuation across the projected plane of galaxy. We then subtract this extinction from the intrinsic luminosity at each pixel, ensuring that the resulting brightness reflects the physical impact of dust absorption and scattering. This process aligns the mock observations of star-forming galaxies with the effects observed in real data.
		
	\subsection{Mock SDSS Galaxy}\label{Sec_method_mock}
	Through our established methodology, we generated galaxy and group images of simulated galaxies at various redshifts. However, unlike real galaxy observations, these images are devoid of instrumental effects and noise. To bridge this gap, we applied PSF convolution, introduced sky background, and simulate noise, aiming to emulate the characteristic image data from the SDSS, following the methodology presented in \cite{deGraaff_2022}. 
	
	In our data processing pipeline for generating mock observation data from TNG100 simulations, PSF convolution plays a pivotal role. Accurate PSF samples are crucial to ensure the fidelity of these mock observations. We adopted the PSF calculation methodology outlined in Appendix A of \cite{Abdurro2021}, tailoring it to our specific dataset. For each galaxy in the \customcite{Nair_2010}{N10} catalog, we constructed an empirical PSF (ePSF) after subtracting the background from the galaxy image. This process involved selecting nearby stars, carefully chosen through visual inspection to exclude those with overlapping features or saturation. We observed that the ePSF shapes varied across these \customcite{Nair_2010}{N10} galaxies, likely due to differing seeing conditions and instrumental effects. To address this variability, we stacked and averaged all ePSFs to create a composite ePSF, which was then used for convolving the TNG100 data and analyzing the morphological parameters of the mock galaxies. The composite PSF aligns well with theoretical expectations and is accurately represented by a double Gaussian profile with a Full Width Half Maximum (FWHM) of approximately 1.5 arcseconds, matching the typical seeing conditions in SDSS.
	
	In preparing our mock SDSS images using TNG100 data, we introduced a constant background with a brightness level set to 20.9 mag arcsec$^{-2}$, mirroring the sky background level. For the simulation of noise, we initiated our process by referencing standard parameters from the SDSS DR9 catalog for the $r$-band, including the detector gain ($G = 4.73 e^{-} ADU^{-1}$), and the conversion factor from counts to fluxes ($nMgyPerCount = 0.0051 nMgy~ADU^{-1}$). The image was then converted to units of $e^{-} pix^{-1}$, simulating the collection of photoelectrons by the detector. We estimated the noise of these photoelectrons by considering a Poisson distribution for each pixel, where the mean value ($\mu$) corresponds to the number of electrons ($N_{e,pix}$) in that pixel. A random value was drawn from this distribution and added as noise to the corresponding pixel. After dividing the image by the gain, and subtracting a constant background, we completed the preparation of the science-ready mock SDSS observation data. The processed images emulating SDSS observations will be referred to as mock observation data throughout the remainder of this text. All the data processing tasks mentioned above were facilitated by the \texttt{Python} package \textit{Photutils} (\citealt{photutils2020}).

	\begin{figure*}[t]
		\centering
		\includegraphics[width=\textwidth]{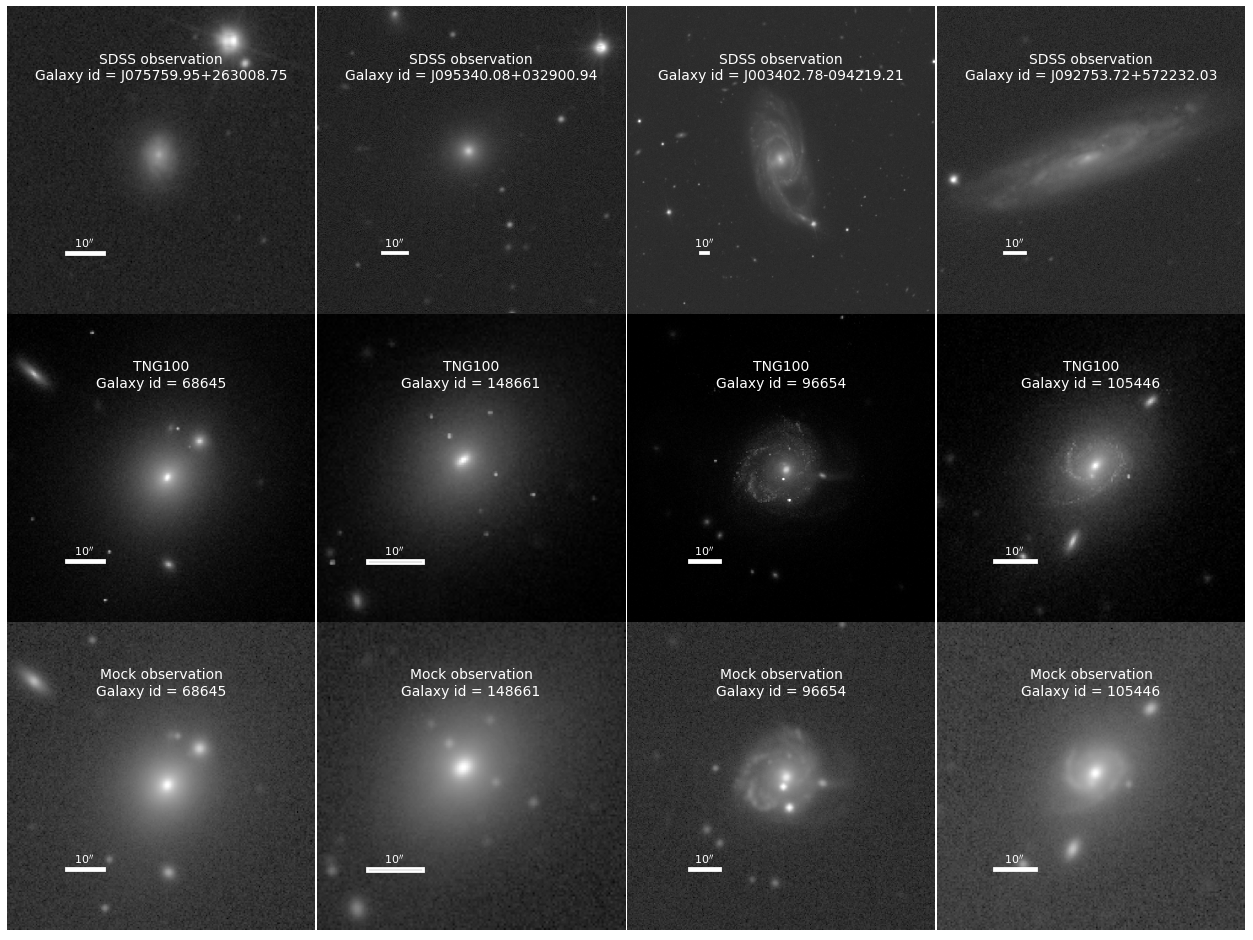}
		\caption{Illustration of SDSS observations, TNG100 data and the resultant of mock observations. The top row displays actual SDSS observational data for selected galaxies, with object IDs (SDSS J2000 object identifier) provided. The middle row exhibits the TNG100 data galaxies in the $x \text{-} y$ projected plane for $z = 0.01$. The bottom row presents the TNG100 data after undergoing our mock observation processing, which includes PSF convolution, background addition, and noise simulation. The left two columns are ETGs, while the right two columns are LTGs. The scale in the bottom-left corner indicates the $10^{\prime\prime}$ scale. }        \label{fig:show_galaxy_sample}
	\end{figure*}
	
	The accuracy of our simulations in mirroring real observations is crucial for confirming the effectiveness of the techniques we use to study galaxy morphology. The comparison depicted in Figure \ref{fig:show_galaxy_sample} illustrates the visual concordance between SDSS data, TNG100 data, and the mock observation data. As a reminder, SDSS data represents the observational data from the \customcite{Nair_2010}{N10} sample (Section \ref{SDSS_data}), TNG100 data refers to the original TNG100 simulation data without accounting for observational conditions (Section \ref{Sec_tl_data}), and mock observation data corresponds to TNG100 data processed to mimic the observational conditions of SDSS (this section). Figure \ref{fig:show_galaxy_sample} shows that the mock observation galaxies appear more blurred than those in the TNG100 data, but resemble those in SDSS. This clearly demonstrates our method's capability to replicate the appearance and features of galaxies observed in the SDSS data.
	
	\subsection{Classify Galaxies via Deep Learning}\label{sec_method_CNN}
	The T-type classification system (\citealt{deVaucouleurs1959,deVaucouleurs1963}) is a widely adopted numerical scheme for quantifying galaxy morphology. Along the Hubble sequence, it offers a continuous sequence of galaxy types, ranging from ETGs with dominant bulges (low T-type values, typically T-type $\leq 0$, including ellipticals and lenticulars) to LTGs with prominent disks (high T-type values, typically T-type $> 0$, encompassing spirals and irregulars). Intermediate T-type values often represent galaxies with transitional morphologies. This system consolidates complex structural features into a single, continuous parameter, enabling objective quantitative analysis (\citealt{Kochanek2001, Conselice2006, Yoon2021}). In this work, we leverage these T-type classifications as the foundation for building our CNN training set, and while our CNN construction aligns with the framework presented in \cite{Domnguez2018} and \cite{Huertas-Company_2019} (hereafter referred to as \customcite{Huertas-Company_2019}{H19}), our application involves slight modifications to adapt the architecture to our specific research objectives.
	
	To this end, we utilize the galaxy images corresponding to the morphological classifications provided in the \customcite{Nair_2010}{N10} catalog\footnote{Galaxy classified as `unknown' (T-type \( = 99 \)) are not included in this work.} as the training set for our CNN models. These images are uniformly resized to \( 128 \times 128 \) pixels, a resolution sufficient to encompass the regions of almost all galaxies in the catalog, and preprocessed by subtracting the median background and normalizing pixel values to their peak values, ensuring a consistent range between 0 and 255. To minimize the impact of sample quality on model performance, we excluded galaxies compromised by contamination, such as those affected by foreground stars, bad pixels, or partial luminosity loss due to edge placement. 
	
	Our CNN model was trained using the \textit{Keras} task in \texttt{Python}. The architecture consists of four convolutional layers with filter sizes of $6 \times 6, 5 \times 5, 4 \times 4$, and $3 \times 3$, using 32, 64, 128, and 128 filters respectively. Each convolutional layer is followed by a dropout layer to prevent overfitting, with rates of 0.5, 0.25, 0.25, and 0.25. After every second and third convolutional layers, a $2 \times 2$ max-pooling layer is added. After the convolutional stages, the feature maps are flattened and passed through a dense layer with 64 neurons. The network ends with a single output neuron with a sigmoid activation, suitable for binary classification.
	
	In this work, the model was compiled using a binary cross-entropy loss function, appropriate for binary classification tasks, with the specified \textit{adam} optimizer. We trained the model with a batch size of 30 over 200 epochs. To enhance the model's performance in this binary classification context, we applied data augmentation techniques during training. This included rotating images up to 45 degrees, allowing horizontal and vertical shifts of up to 5\%, zoom adjustments between 0.75 to 1.3, and both horizontal and vertical image flipping. Training was conducted on this augmented dataset, with concurrent validation on a separate dataset. Our CNN architecture incorporates design choices and parameter settings that have been successfully applied in previous studies (\citealt{Huertas-Company_2019, Dominguez_2022, Walmsley2023}).
	
	We define three hierarchical models based on T-type values: Model 1 classifies galaxies into ETGs (\( \textnormal{T-type} \leq 0 \)) and LTGs (\( \textnormal{T-type} > 0 \)), with a balanced training set of 5,645 galaxies for each type. Model 2, trained on ETGs identified by Model 1, further distinguishes elliptical galaxies (\( \textnormal{T-type} \leq -3 \)) and S0/a galaxies (\( -3 < \textnormal{T-type} \leq 0 \)), using 2,057 training galaxies per class. Model 3, trained on LTGs identified by Model 1, separates Sab/Sb galaxies (\( 1 \leq \textnormal{T-type} < 4 \)) from Sc/Sd/irregular galaxies (\( \textnormal{T-type} \geq 4 \)), with 2,664 training galaxies per class. To better capture subtle morphological differences, we adjusted the CNN architectures for Model 2 and Model 3 by removing max-pooling layers in the first and fourth convolutional layers, reducing dropout rates to 0.1 (after the second convolutional layer) and 0.2 (before the dense layer), and introducing L2 regularization (factor of 0.0001) in the dense layers. These modifications enhance the sensitivity of models to nuanced features, improving classification reliability for closely related galaxy types.
	
	To evaluate our CNN models, we first assessed their performance during training using validation loss, accuracy, and F1-score. Validation loss quantifies the discrepancy between model predictions and ground truth, with lower values indicating better performance. Accuracy measures the fraction of correctly classified galaxies, while F1-score, the harmonic mean of precision and recall, balances false positives and false negatives. Model 1 achieved a loss of 0.13, accuracy of 0.95, and F1 of 0.95; Model 2 achieved a loss of 0.36, accuracy of 0.85, and F1 of 0.85; Model 3 achieved a loss of 0.50, accuracy of 0.77, and F1 of 0.77.
	
	Practical validation was performed using two independent catalogs. For Model 1, we utilized the catalog from \cite{Schawinski2010}, based on the Galaxy Zoo project (\citealt{Lintott2008,Lintott2011}), excluding galaxies labeled as `merger' and `indeterminate'. Galaxies were randomly selected from three stellar mass bins: \(\log(M_*/M_{\odot})\) in [9,10], [10,11], and [11,12], with 600 ETGs and 600 LTGs chosen from each bin\footnote{The reason for selecting 600 galaxies from each bin was constrained by the catalog, which contained just over 600 ETGs within the [9,10] stellar mass range.}. Model 1 achieved a remarkable accuracy of 94.75\% on this sample of 3,600 galaxies, confirming its ability to distinguish ETGs from LTGs. However, the \(\sim\)5.25\% error rate partly arises from an uneven stellar mass distribution in the N10 catalog training set, where low-mass galaxies (\(\log(M_*/M_{\odot})\) in [8,9]) are mostly LTGs and high-mass ones ([11,12]) are mostly ETGs, leading to misclassification rates of \(\sim\)10.00\% for these low-mass ETGs and 6.78\% for these high-mass LTGs. Since the \cite{Schawinski2010} catalog only provides classifications at the ETG and LTG levels, it lacks the detailed morphological subtypes needed to evaluate Models 2 and 3. Therefore, we employed the Formes Idéalisées de Galaxies en Imagerie (EFIGI) catalog (\citealt{Baillard2011}), which provides finer morphological classifications, including T-type values for each galaxy. Model 2 achieved an accuracy of 84.93\% in distinguishing elliptical galaxies from S0/a galaxies, while Model 3 reached 83.38\% in classifying Sab/Sb galaxies versus Sc/Sd/irregular galaxies. For comparison, Model 1 was also evaluated on the EFIGI catalog, achieving an accuracy of 91.99\%. These results demonstrate the robustness of our CNN models in capturing morphological characteristics across diverse galaxy populations. However, the lower accuracies of Models 2 and 3 compared to Model 1 may stem from a significant reduction in training sample size.
	
	\subsection{Quantification of Galaxy Morphology}\label{sec_method_morphology}
	
	In this study, we focus on quantifying galaxy properties through three main approaches: non-parametric statistics, \sersic{} function fitting, and aperture photometry. This analysis covers a comprehensive sample of galaxies, encompassing both mock observational galaxies and actual galaxies observed in SDSS. 
	
	Galaxy segmentation maps are crucial for the analysis of galaxy morphology because they delineate the spatial extents of galaxies in images, effectively separating regions where galaxy signals dominate from those where background noise prevails. Additionally, they can mask the influence of nearby bright sources on the analysis of a galaxy's own properties and provide a constrained set of guesses for fitting parameters. We first convolved both the mock and SDSS images with a Gaussian kernel having a FWHM of $1^{\prime\prime}.5$ \footnote{This convolution is only for segmentation purposes.}. For segmentation, each galaxy is initially identified by marking pixels that are more than 1.5 times the background uncertainty\footnote{Background uncertainty is defined as the median value of pixels after background subtraction.} and comprise at least five pixels as part of the source. During deblending, we attempt to divide the brightness of each source into 20 equal parts, requiring at least 20 pixels and a total pixel brightness that accounts for at least 10\% of the peak brightness of the source to be considered as a separate source\footnote{Set by parameter \textit{contrast} and \textit{nlevels} in \textit{Photutils.deblend\_sources}.}. This step aims to uniquely label different sources while ensuring that a single galaxy is not fragmented into multiple pieces. However, these parameters may not be suitable for all samples, especially spiral galaxies, which are prone to being oversampled, resulting in the galaxy being divided into multiple parts, or undersampled, leading to multiple sources being assigned the same label. In such cases, manual adjustment of the parameters is necessary to ensure that each independent source is appropriately labeled, while the label out of the target galaxy will serve as the mask image.
	
	To analyze the morphology of galaxies in our dataset, we utilized the \textit{Statmorph} package in \texttt{Python} to derive a range of non-parametric morphological statistics for each galaxy. The segmentation maps required for running \textit{Statmorph} have been discussed previously. Furthermore, utilizing an accurate PSF image is crucial for the reliable calculation of several morphological parameters. Therefore, we specified the appropriate PSF for each dataset during this analysis. For the morphological fitting of the mock observation galaxies, the composite ePSF described in Section \ref{Sec_method_mock} was used. Conversely, for the morphological fitting of the actual SDSS galaxies, the specific ePSF for each galaxy was used in the fitting process to accurately account for the PSF variations caused by differing seeing conditions during each observation. The Gini--$M_{20}$ classification system (\citealt{Lotz2004,Zamojski2007,Snyder2015}) is a widely recognized method for quantifying galaxy morphologies. Alongside, the Concentration-Asymmetry-Smoothness (CAS) system serves as another foundational set of non-parametric morphological indicators (\citealt{Conselice2003,Mortlock2013,Whitney2021}). For an in-depth understanding of these non-parametric morphologies, \cite{Rodriguez-Gomez_2019} (hereafter \customcite{Rodriguez-Gomez_2019}{R19}) provides comprehensive computation details. This section offers a succinct overview of the measurement approaches for each statistic.
	\paragraph{Gini Coefficient (GINI)}
	The Gini coefficient quantifies the inequality among values of a frequency distribution, such as levels of light within a galaxy image. It is defined as:
	\begin{equation}
		G = \frac{1}{\bar{X} n (n-1)} \sum_{i=1}^{n} (2i - n - 1) x_i,
	\end{equation}
	where $n$ is the number of pixels, $x_i$ is the pixel value after sorting them in increasing order, and $\bar{X}$ is the mean pixel value. When all fluxes are concentrated in a single pixel, a value of $G = 1$ is obtained, while a uniform brightness distribution produces $G = 0$.
	\paragraph{$M_{20}$ Coefficient}
	The $M_{20}$ coefficient measures the spatial distribution of the brightest 20\% of a galaxy's pixels. It is calculated as:
	\begin{equation}
		M_{20} = \log_{10}\left(\frac{\sum_i M_i}{M_{\text{tot}}}\right), \text{ while } \sum_i f_i < 0.2f_{\text{tot}},
	\end{equation}
	where $M_i = f_i \times [(x_i - x_c)^2 + (y_i - y_c)^2]$, $f_i$ is the flux in pixel $i$, and $f_{\text{tot}}$ is the total flux of the galaxy. $x_c, y_c$ are the coordinates of the galaxy's center. While the Gini coefficient provides a measure of the overall distribution of pixel values, indicating how light is shared across the galaxy, the $M_{20}$ coefficient specifically targets the spatial organization of the galaxy's brightest regions. This contrast allows $M_{20}$ to offer insights into the galaxy's structural composition and central concentration that Gini alone may not fully capture.
	\paragraph{Concentration (C)}
	The concentration index compares the brightness within a central aperture to the overall brightness of the galaxy, reflecting its light concentration. It is expressed as:
	\begin{equation}
		C = 5 \times \log_{10}\left(\frac{r_{80}}{r_{20}}\right),
	\end{equation}
	where $r_{80}$ and $r_{20}$ are the radii containing 80\% and 20\% of the total galaxy light, respectively. While $r_{20}$ and $r_{80}$ are the conventional radii used to define concentration index $C$ here, it's worth noting that some studies may choose alternative radii definitions (e.g., using $r_{90}$ and $r_{50}$ to define $C_{95}$). 
	\paragraph{Asymmetry (A)}
	The asymmetry index assesses the degree to which a galaxy's light distribution is rotationally symmetric. It is defined by:
	\begin{equation}
		A = \frac{\sum_{i,j} | I_{ij} - I_{ij}^\prime |}{\sum_{i,j} | I_{ij} |} - A_{\text{bgr}},
	\end{equation}
	where $I_{ij}$ and $I_{ij}^\prime$ are the original and 180-degree rotated image pixels, respectively, and $A_{\text{bgr}}$ is the average asymmetry of the background.
	\paragraph{Smoothness (S)}
	The smoothness index, or clumpiness, indicates the presence of small-scale structures within a galaxy by comparing the galaxy image to a smoothed version of itself. It is given by:
	\begin{equation}
		S = \frac{\sum_{i,j}  I_{ij} - I_{ij}^S }{\sum_{i,j}  I_{ij} } - S_{\text{bgr}},
	\end{equation}
	where $I_{ij}^S$ represents the smoothed image, achieved by convolving the original image with a box of a specific size, and $S_{\text{bgr}}$  is average smoothness of the background.
	
	Beyond non-parametric statistics, fitting the light distribution of galaxies with the \sersic{} function is a prevalent method for analyzing galaxy morphology. We conducted fittings using the widely recognized software \texttt{GALFIT}(\citealt{Peng2002,Peng2010}), following the methodology outlined in \cite{Gong2023}. \texttt{GALFIT} requires input images including the PSF image, a mask image, and a sigma image. The construction of the PSF image has been discussed in Section \ref{Sec_method_mock}, and the mask image in Section \ref{sec_method_morphology}. For the sigma image creation, galaxy images without background subtraction were converted to electron units to account for Poisson noise for each pixel. For SDSS data, additional consideration for the read noise is included, and the final sigma image is then converted to the same units as the galaxy image. The \sersic{} profile is described by the equation:
	\begin{equation}
		I(r) = I_e \exp \left\{ -\kappa \left[ \left( \frac{r}{R_e} \right)^{1/n} - 1 \right] \right\},
	\end{equation}
	where $I(r)$ signifies the intensity at radius $r$, $I_e$ is the intensity at the effective radius $R_e$, $n$ is the \sersic{} index, and $\kappa$ is a coefficient associated with $n$. For single \sersic{} profile fitting, the \textit{data\_properties} function from \textit{Photutils} task  employed to extract geometric properties of the segmentation of target galaxy, such as centroid coordinates, semi-major and semi-minor axes, and position angle. These properties provided initial estimates for the \texttt{GALFIT} executions. $n$ was determined based on the galaxy classification from our CNN model: $n = 1$ for LTG and $n = 3$ for ETG. The galaxy's magnitude was calculated using Equation \ref{equ:mag}. To enhance the fitting accuracy, we executed three iterations of \texttt{GALFIT}, with the initial parameters for the second and third iterations derived from the best-fit results of the preceding iteration.
	
	In pursuit of more accurate defining galaxy apertures, isophotal fitting was performed for each galaxy. We adopted the method outlined by \cite{Gong2023}, which simplifies the approach initially proposed by \cite{Li_2011}. Utilizing the \textit{Ellipse} function from the \textit{Photutils} task, along with previously discussed galaxy segmentation maps and their geometric properties, we fitted the isophotal profiles of the galaxies. The aperture for each galaxy was defined by the isophote corresponding to a surface brightness of 26.5 mag arcsec$^{-2}$. For mock observation galaxies, this elliptical aperture, in conjunction with the stellar mass maps, was further employed to measure the total stellar mass and SFR, while the age and metallicity estimations of stellar populations required stellar mass-weighting.
	
	\section{Result}\label{Sec_result}
	
	During the fitting and computation processes, certain data points inevitably yield warnings or physically implausible outputs. To ensure the integrity of our analysis, it was necessary to exclude these anomalous results. For non-parametric statistics, the \textit{Statmorph} package provides a `flag' parameter that indicates the quality of each fit; we retained only those galaxies with a `flag' value of 0, which means a good fit. In the case of \sersic{} fitting, results exhibiting a \sersic{} index greater than 10 were discarded, as were those galaxies which isophote fitting could not be successfully executed, regardless of manual adjustments. Furthermore, considering the observational limits of the SDSS and to maintain consistency with the brightness distribution of the \customcite{Nair_2010}{N10} catalog, we excluded samples with $g$-band absolute magnitudes fainter than -16.8. Our final catalog comprises 34,118 mock observation galaxies, each with 43 attributes, representing a substantial and comprehensive dataset for further analysis. A subset of these galaxies, specifically nine randomly selected examples, is detailed in Table \ref{tab:catalog}, and the explanations of all parameters are all in the appendix (this table is available in its entirety in machine-readable form in the online article.).  
	
	\subsection{Morphologic Parameters}
	
	\begin{figure*}[t]
		\centering
		\includegraphics[width=\textwidth]{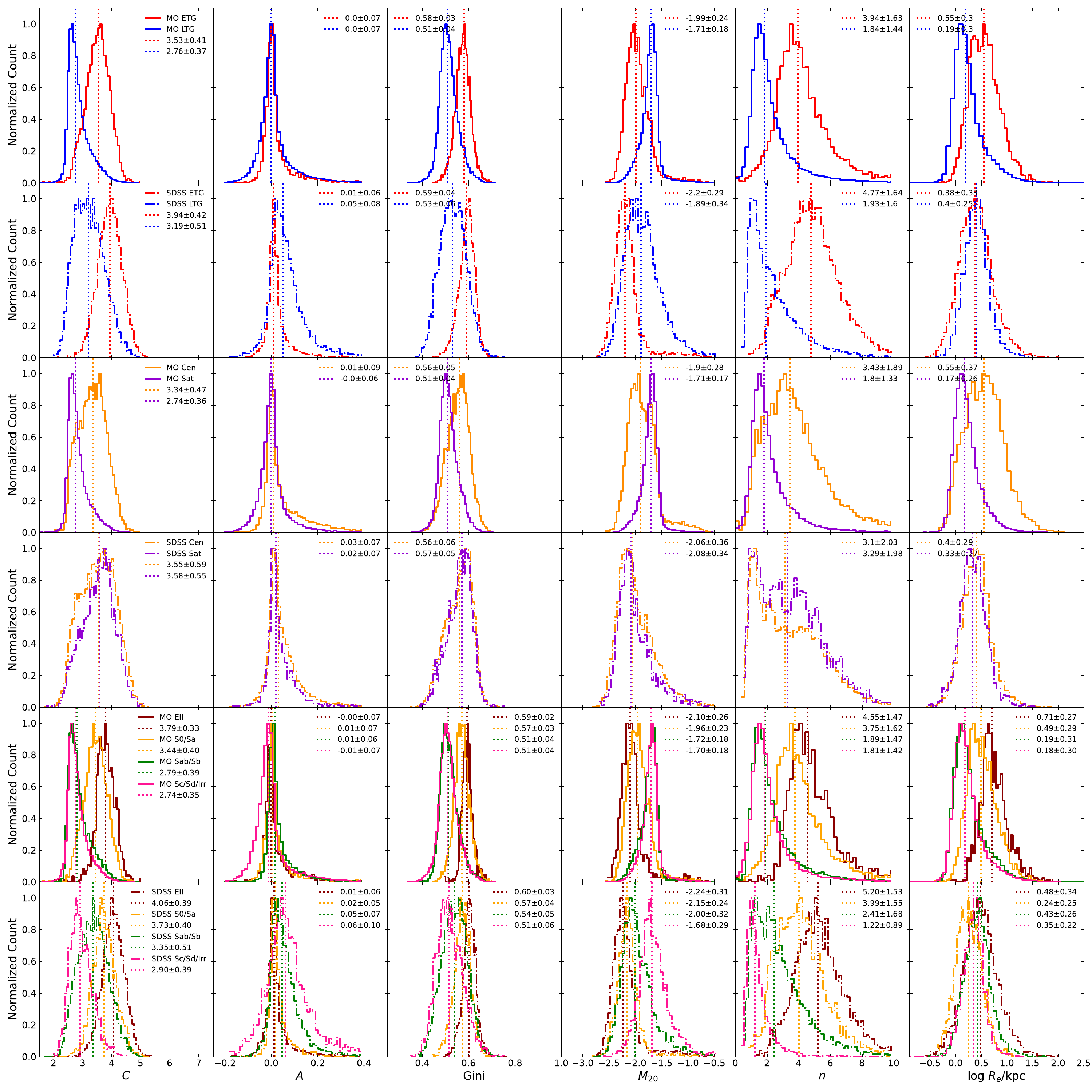}
		\caption{Normalized histograms illustrating the comparison between mock observation (labeled as ``MO" in the figure) and SDSS data for various galaxy morphological parameters. Each subplot represents a different parameter: $C$, $A$, Gini, $M_{20}$, $n$, and $R_e$, arranged from left to right in columns. Solid lines represent mock observation data, while dash-dotted lines represent SDSS data. Dotted lines indicate the median values, with error denoting the standard deviation. Red lines denote ETGs, blue lines represent LTGs, dark orange lines represent central galaxies (labeled as ``Cen" in the figure), and dark purple lines represent satellite galaxies(labeled as ``Sat" in the figure). The figure also shows four specific morphological types: elliptical galaxies (labeled as ``Ell'' in the figure) with deep red lines, S0/a galaxies with orange lines, Sab/Sb galaxies with green lines, and Sc/Sd/Irr galaxies with deep pink lines. All histograms are normalized to a maximum value of 1.}
		\label{fig:allparameters_hist}
	\end{figure*}
	
	\begin{figure*}[t]
		\centering
		\includegraphics[width=\textwidth]{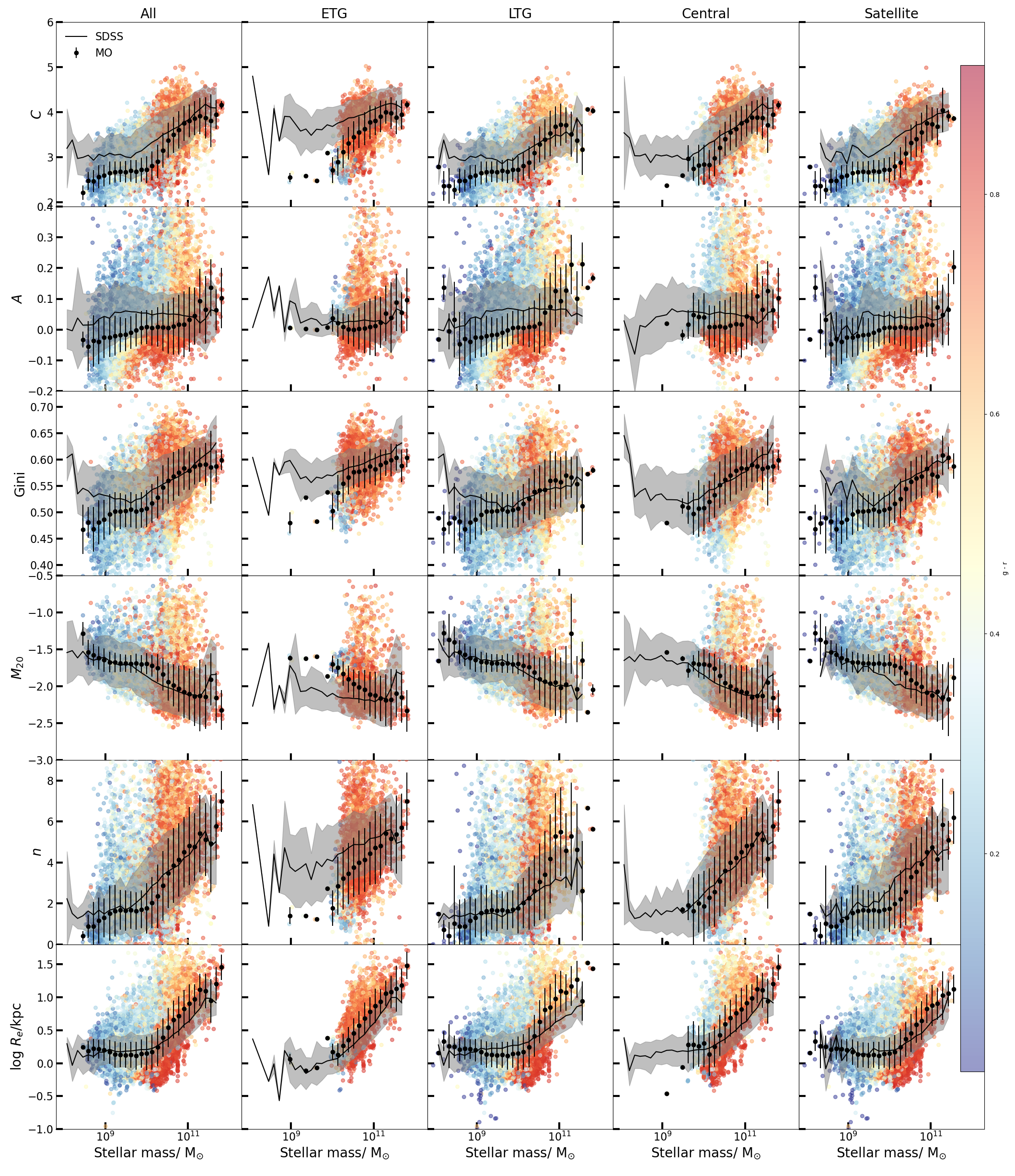}
		\caption{The relationship between various galaxy morphological parameters and stellar mass is shown for different galaxy samples: all galaxies, ETG, LTG, central galaxies, and satellite galaxies. From top to bottom, the rows represent $C$, $A$, Gini, $M_{20}$, $n$, and $R_e$. In each panel, data points indicate mock observation (MO) results, with colors representing the $g-r$ color (redder points for higher $g-r$, bluer points for lower $g-r$). Black points show the median values within each stellar mass bin, with error bars denoting the standard deviation. The black solid line represents the SDSS results, with the grey shaded area illustrating the standard deviation as error.}
		\label{fig:allparameters_vs_mass}
	\end{figure*}
	
	\begin{figure*}[t]
		\centering
		\includegraphics[width=\textwidth]{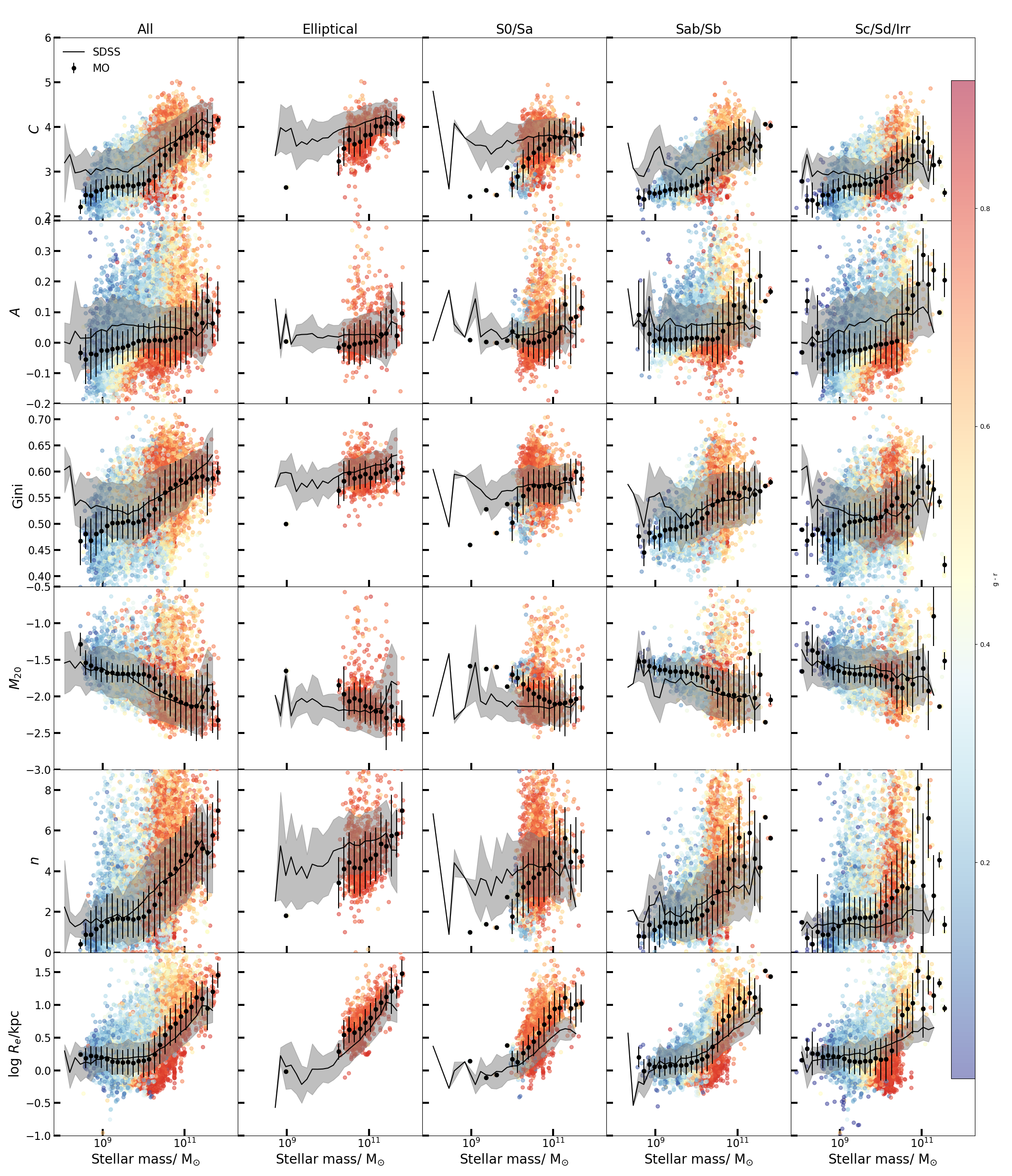}
		\caption{The same as Figure \ref{fig:allparameters_vs_mass}, but showing the relationship between various galaxy morphological parameters and stellar mass for different galaxy Type: all galaxies, elliptical galaxies, S0/a galaxies, Sab/Sb galaxies, and Sc/Sd/Irr galaxies. 
		}
		\label{fig:allparameters_withtype_vs_mass}
	\end{figure*}
	
	Figure \ref{fig:allparameters_hist} presents histograms of morphological parameters for different galaxy types within the mock observation and SDSS sub-samples, showing the similarities and differences in their morphological properties. The figure demonstrates that the mock observations can robustly reproduce the morphological differences between ETGs and LTGs, showing that ETGs exhibit higher concentration indices (including $C$ and $n$) compared to LTGs. This characteristic is also evident in the SDSS sub-sample. However, the concentration parameters of all galaxies in the mock observations are systematically lower than those of their SDSS counterparts. Additionally, the median $R_e$ of ETGs in the mock observations is higher than that of LTGs, a trend not as clearly observed in the SDSS data. 
	
	For central and satellite galaxies, we adopt a classification scheme consistent with that used in the SDSS sample. According to \customcite{Nair_2010}{N10}, this classification is based on the group catalog from \cite{Yang2007}, which identifies whether a galaxy is the most massive or the most luminous member of its group. Following the same criteria, we classify the most massive galaxy in each group as the central galaxy in our mock observations, with all other members designated as satellite galaxies. Under this classification, the mock observations from TNG100 suggest that central galaxies are predominantly ETGs, while satellites are mainly LTGs. In contrast, the SDSS data analyzed in this work do not show a significant morphological difference between central and satellite galaxies. This latter finding is somewhat counterintuitive, as central and satellite galaxies occupy distinct environments that are widely expected to influence their formation and evolutionary pathways, often resulting in differing morphological characteristics (\citealt{Weinmann2006,van2008,Tempel2011,Wilman2012,Chen2024}). Our investigation indicates that a reason for the lack of a clear morphological distinction in our SDSS analysis is the contamination of the central galaxy sample by isolated field galaxies. This contamination arises because the adopted SDSS group catalog classifies some isolated systems as single-member groups with a central galaxy. Since isolated galaxies evolve in different environments compared to true centrals within groups or clusters, their inclusion likely biases the results for the central population. Furthermore, while generally rare, the situation where the most massive galaxy in a system is not located at the point of minimum gravitational potential could also introduce minor uncertainties.
	
	The bottom two rows of Figure \ref{fig:allparameters_hist} present a comparison of morphological parameters between mock observation galaxies and SDSS galaxies across different morphological types. While the Gini coefficient ($G$) exhibits the highest agreement across all morphological types, followed by the $M_{20}$ and $A$ parameters, the concentration parameters in mock observations are systematically lower than those in SDSS counterparts. However, all parameters remain statistically consistent within their error bars. Notably, both datasets display a clear morphological gradient with morphological type: galaxies progress from late-type spirals to intermediate types and finally to ellipticals, showing a monotonic increase in structural compactness. Furthermore, the analysis indicates that the morphological parameters distinguishing Sab/Sb from Sc/Sd/Irr galaxies show significantly less variation in the TNG100 mocks compared to SDSS (see e.g., bottom two rows). This apparent similarity between these LTG subtypes in the simulation likely stems from its limited resolution, which can hinder the differentiation based on fine structural details that are observable in higher-resolution SDSS data.
	
	Figure \ref{fig:allparameters_vs_mass} presents the dependence of various morphological parameters on stellar mass ($M_*$) for different galaxy subsamples. While Figure \ref{fig:allparameters_hist} shows the overall distribution of morphological parameters across different galaxy types, Figure \ref{fig:allparameters_vs_mass} illustrates these parameters as a function of stellar mass in different bins, revealing that the observed differences in morphological properties are less pronounced when considering within specific mass bins. A clear trend is observed where the $g-r$ color becomes redder with increasing stellar mass. Specifically, ETGs are predominantly red, while LTGs exhibit a mix of both red and blue galaxies. The comparisons between the SDSS (black solid lines) and the mock observations (black solid points) for parameters such as $A$, Gini, $M_{20}$, and $R_e$ show remarkable agreement within the errors. This consistency suggests that the methods proposed in this study effectively reproduce the morphological characteristics of simulated galaxies in line with observed galaxies. Notably, $R_e$ displays the highest level of consistency, with the mass-size relation from the mock observations closely matching the SDSS results \citep[see also Fig.4 in][]{Tang_2021}. However, at both the high and low mass ends, there is a noticeable increase in scatter in these parameters, which may lead to deviations between the mock observations and SDSS results. This increased scatter is likely due to the smaller number of galaxies in these mass bins.
	
	In addition to the overall consistency, Figure \ref{fig:allparameters_vs_mass} reveals significant discrepancies in morphological parameters between mock observation galaxies and their SDSS counterparts, particularly at the low-mass end across all subsamples. Across all subsamples, the $C$ parameter increasingly deviates as stellar mass decreases, with a more pronounced offset in mock observation galaxies at lower masses. Similarly, ETGs exhibit a growing discrepancy from SDSS values in parameters such as $C$, $M_{20}$, and $n$ with decreasing stellar mass. For central and satellite galaxies, the morphological parameters of central galaxies closely align with SDSS galaxies, indicating high consistency. However, satellite galaxies show significant deviations, especially in the $C$ parameter below $10^{10} \, M_\odot$. These discrepancies are likely attributed to the resolution limitations of the TNG100 simulation, such as the impact of softening length on galaxy surface brightness profiles (\citealt{deAraujoFerreira2025}). Notably, for stellar masses below $10^{10} \, M_\odot$, the number of mock observed ETGs and central galaxies drops sharply, and these galaxies display significantly lower $C$ and $n$ values compared to their SDSS counterparts. 
	
	Figure \ref{fig:allparameters_withtype_vs_mass} presents the morphological parameter distributions as a function of stellar mass for different morphological subtypes in both the mock observation and SDSS datasets. Similar to Figure \ref{fig:allparameters_vs_mass}, this figure follows the same format but focuses on the differences across more detailed galaxy classifications. Among all types, elliptical galaxies exhibit the highest consistency between mock observations and SDSS counterparts, with morphological parameters closely matching within the error bars across the entire mass range. For other galaxy types, minor differences exist, but they remain within the error range. As seen in Figure \ref{fig:allparameters_vs_mass}, the limited sample size at both the high-mass and low-mass ends leads to increased scatter. Notably, even after subdividing galaxies by morphology, deviations in structural parameters, particularly concentration index \( C \) and \( n \), become more pronounced for galaxies with stellar masses below \( 10^{10} M_{\odot} \), compared to other parameters.

	\subsection{The Proportion of ETGs and LTGs}
	
	\begin{figure*}[t]
		\centering
		\includegraphics[width=\textwidth]{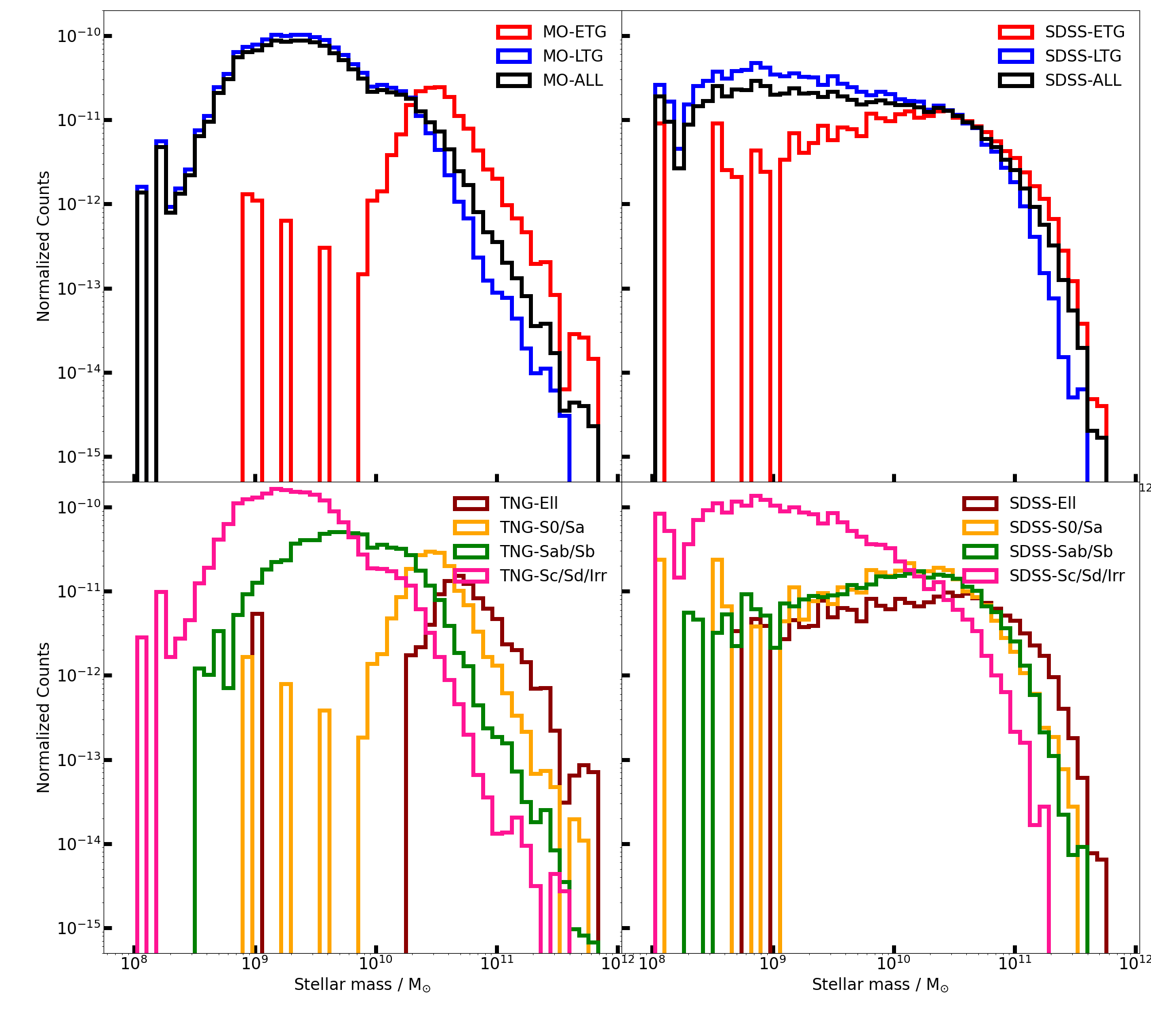}
		\caption{Stellar mass histograms for mock observation data (left column) and SDSS data (right column). The top row shows ETGs (in red) and LTGs (in blue), and the combined sample of all galaxies (black). The bottom row is separated into morphological subsamples, including Ellipticals (darkred), S0/Sa (orange), Sab/Sb (green), and Sc/Sd/Irr (deeppink). Each data is normalized such that the integral area equals 1.}
		\label{fig:mass_hist}
	\end{figure*}
	
	The variation in galaxy counts across different stellar mass bins is depicted in Figure \ref{fig:mass_hist}. For dwarf galaxies ($M_*/M_{\odot} < 10^{10}$), the number of LTGs in the mock observations significantly exceeds that observed in the SDSS, while conversely, the count of ETGs in the mock observations is notably lower than that in the SDSS. Specifically, ETGs constitute only 14.00\% of the total galaxy population in the mock observations, compared to 42.52\% in the SDSS. Furthermore, when considering galaxies with stellar masses below $10^{10} M_{\odot}$, the proportion of ETGs in the mock observations further decreases to a mere 0.05\%, whereas in the SDSS, at least 19.79\% of galaxies are ETGs. When examining the stellar mass distribution of galaxies with specific morphological classifications, significant differences emerge between the mock observation sample and the SDSS dataset. Notably, the mock observation sample exhibits a pronounced gradient in the stellar mass distribution across different morphological types, a feature that is less apparent in the SDSS dataset. In particular, elliptical galaxies in the mock observation sample are predominantly found in the highest stellar mass range. Conversely, late-type spiral galaxies are more concentrated in the low-mass regime. For S0/Sa galaxies, the mock observation sample contains virtually no objects with stellar masses below $10^{10} \, M_\odot$, whereas the SDSS dataset includes a substantial number of such low-mass systems. Regarding Sab and Sb galaxies, the mock observation results indicate that these systems are primarily concentrated within the $10^9 - 10^{10.5} \, M_\odot$ range. In contrast, the corresponding SDSS counterparts exhibit a more uniform distribution across this mass interval. It is pertinent to remind readers that the SDSS data utilized in this work follows the \citet{Nair_2010} catalog, which was aimed at providing a highly detailed galaxy classification sample and thus might be subject to selection biases. On the simulation side, the substantial volume of TNG100 ensures that statistical uncertainties arising from cosmic variance are expected to be minor relative to typical observational uncertainties, a point established for its progenitor simulation by \citet{Genel2014}. A more in-depth analysis will be detailed in the Section \ref{sec_lack_of_etg}. 
	
	\section{Discussion}\label{Sec_disscussion}
	
	\subsection{The Influence of Observation Conditions on Morphological Classification}
	
	It is well known that astronomical observations are influenced by observational conditions, such as the effects of the PSF, sky background, and instrumental noise, whereas numerical simulation data remain unaffected by these factors. Consequently, we should examine whether these observational conditions could potentially mislead the CNN in classifying mock observation galaxy images. Some studies have demonstrated that the accuracy of predictions made by supervised CNNs on galaxy morphology is highly sensitive to the data characteristics of the training sample, such as resolution (\citealt{Cabrera2018,Dodge7498955,Medina2023}). Attempts were made to predict the TNG100 sample with our previously constructed CNN models (Models 1 to 3, see Section \ref{sec_method_CNN} for details), yet it became evident that these models struggled to capture the characteristics of galaxies not subjected to mock observations. Therefore, we developed an additional CNN model, Model 4, specifically for the TNG100 dataset (referred to as TNG100 galaxies; see Section \ref{Sec_method_mock} for definition and Figure \ref{fig:show_galaxy_sample} for a visualization of the sample), which was trained on the original projected images without mock observational effects.
	
	To train Model 4 specifically for the TNG100 dataset, we constructed a training set by selecting galaxies from the TNG100 sample at redshift \( z=0.01 \) in the \( x\text{-}y \) projected plane. This selection was guided by the morphological classification results obtained from the CNN trained on mock observation images. Specifically, we included all ETGs and LTGs whose \( n \) and \( C \) fell within the error regions depicted in Figure \ref{fig:allparameters_vs_mass}. To ensure the reliability of this training set, we performed a visual inspection of these galaxies in their original, unprocessed TNG100 images (before applying mock observation effects) to eliminate apparent misclassifications. This newly developed Model 4 was subsequently used to classify the remaining galaxies (those TNG100 galaxies not included in the training set for Model 4) within the TNG100 dataset into ETGs and LTGs. For the galaxies within the training set itself, their classification was directly adopted from the classification assigned during the visual inspection process.
	
	\begin{figure}[t]
		\centering
		\includegraphics[width=0.5\textwidth]{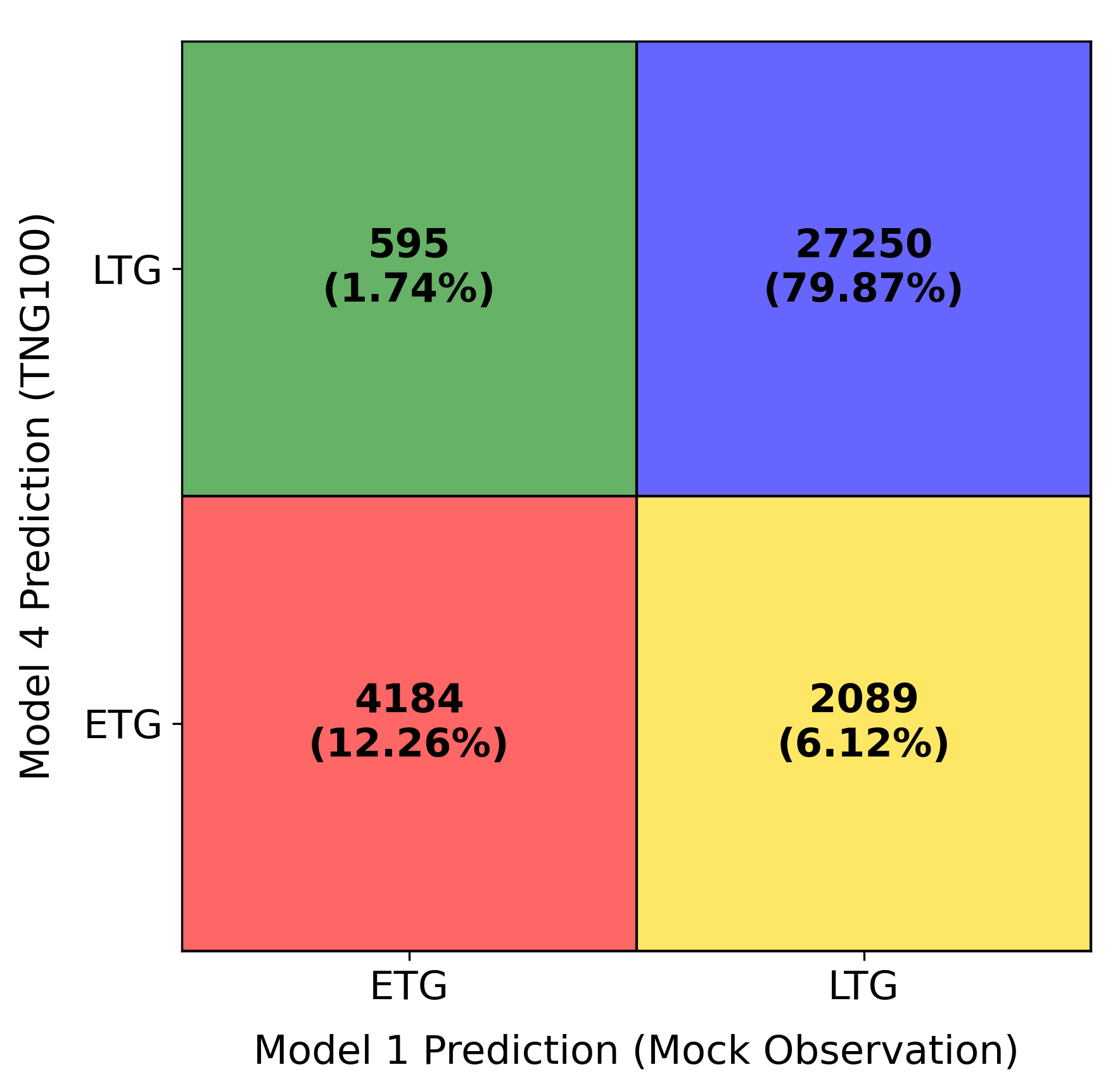}
		\caption{Visualization of the impact of simulated observational effects on galaxy classification via a confusion matrix. This matrix compares the morphological classifications of the same galaxies obtained using Model 1 (applied to the mock observation data) and Model 4 (applied to the original TNG100 data without observational effects). Each cell shows the number count and percentage of galaxies falling into that specific classification combination: red indicates galaxies classified as ETGs by both models; green indicates classification as ETG by Model 1 but LTG by Model 4; yellow indicates LTG by Model 1 but ETG by Model 4; and blue indicates galaxies classified as LTGs by both models.}
		\label{fig:moandtng}
	\end{figure}
	
	Figure \ref{fig:moandtng} presents a confusion matrix comparing the ETG/LTG classifications assigned by Model 1 (on mock observation data) versus Model 4 (on original TNG100 data) for the same galaxies, allowing one to visualize the impact of simulated observational effects. The cells representing consistent classifications (in red and blue) demonstrate that the assigned morphology remains unchanged for the vast majority of galaxies (92.13\%) when comparing the results obtained with and without simulated observational effects. This indicates a high degree of classification stability for most objects. Our analysis therefore focuses on the galaxies with inconsistent classifications (in green and yellow), which constitute the remaining 7.87\% of the sample. To understand why these classification disagreements occur, we examined the structural parameters ($n$ and $C$, derived from the mock data) associated with these specific subsets. Specifically, 1.74\% of the total sample consists of galaxies classified as ETGs by Model 1 but reclassified as LTGs by Model 4 when observational effects are removed (green cell). Analysis of their structural parameters reveals intermediate concentration values ($n \approx 2.77 \pm 1.39$, $C \approx 3.13 \pm 0.37$), suggesting these might be galaxies with transitional morphologies whose inherent structural complexity makes their classification sensitive to observational effects. Conversely, Model 4 reclassifies another fraction (6.12\% of the total sample) of mock observation LTGs (classified by Model 1) as ETGs when observational effects are removed (yellow cell). These galaxies tend to have higher concentrations ($n \approx 3.57 \pm 2.05$, $C \approx 3.27 \pm 0.48$), overlapping significantly with the parameter range of consistently classified ETGs (red cell), potentially indicating a classification bias for some early-type systems under observational conditions. It is worth noting that adopting the classification results from the TNG100 dataset increases the overall proportion of ETGs from the 14.00\% determined by mock observation classifications to 18.39\%. In summary, Figure \ref{fig:moandtng} demonstrates the general robustness of the classification model against observational effects, reliably identifying the morphology for most galaxies. The observed classification discrepancies primarily stem from challenges in classifying galaxies with ambiguous or transitional structures, combined with classification uncertainties introduced by the simulated observational effects.
	
	It is important to clarify the role of Model 4. Its main purpose was to provide a reference classification on the TNG100 data to help quantify the impact of these effects when comparing with Model 1. However, we acknowledge limitations in how Model 4 was constructed. The initial selection of its training sample relied on morphological parameters ($n$ and $C$) derived from mock observations. This selection step, while intended to find galaxies with relatively distinct morphologies for training, likely excluded certain types of galaxies, potentially leading to a training set that may not fully represent the diversity of all galaxy morphologies present in TNG100. For instance, LTGs with large bulges (high $n$) might have been systematically omitted. Despite these limitations, Model 4 still serves as a useful relative benchmark. Comparing classifications from Model 1 and Model 4 allows one to identify the population of galaxies whose assigned class is sensitive to the inclusion of observational effects. Importantly, this demonstrates that observational effects do not only affect ambiguous, borderline cases; even galaxies expected to have relatively clear and distinct morphological features can end up being classified differently solely due to the impact of observational realism.
	
	Considering the inherent probability of errors in the CNN model prediction itself, which has been quantitatively evaluated in Section \ref{sec_method_CNN} using an external catalog to assess the model’s baseline classification uncertainty, the influence of observational conditions on the CNN's accuracy presented here should be regarded as an upper limit. Furthermore, this effect is likely to increase with higher redshift observations, and a challenge not exclusive to machine learning. Studies have also indicated that, visual classifications are susceptible to misjudgments due to observational constraints (\citealt{Cabrera2018,Bamford2009,Willett2017}). The advantage of unsupervised machine learning lies in its independence from training on labeled datasets, presenting a potential avenue to circumvent the challenges posed by observational conditions. However, it is important to acknowledge that unsupervised learning models often entail higher computational and time resources, and their classification result may deviate from the correct physical interpretations. Nevertheless, our methodology, which involves utilizing numerical simulation data to create mock observational images for a perticular survey, offers a novel framework for assessing how observational conditions affect the accuracy of morphological classifications. 
	
	\subsection{The Impact of Convolutional Kernels on Morphological Parameters}
	
\begin{figure*}[t]
		\centering
		\includegraphics[width=\textwidth]{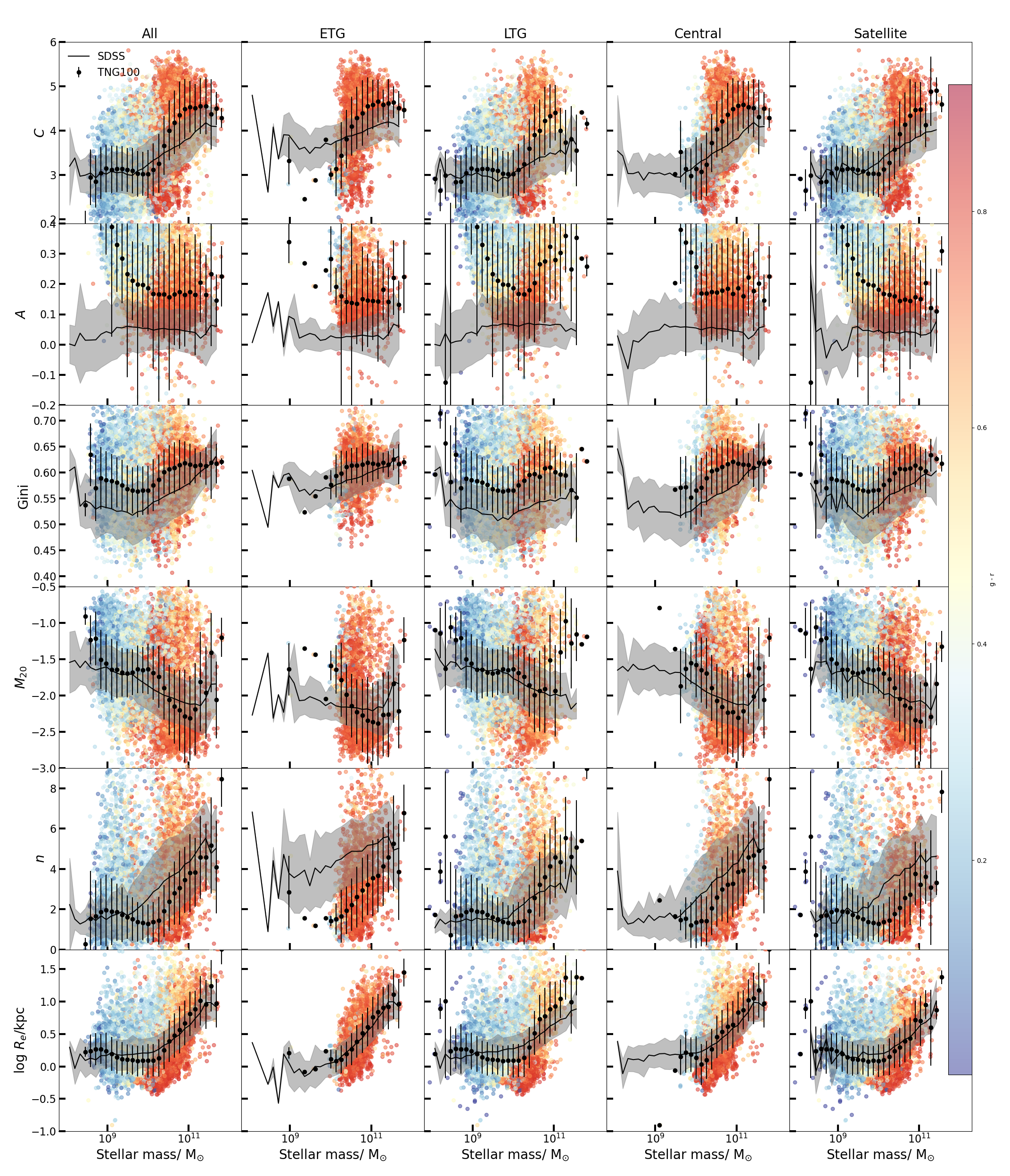}
		\caption{The same as Figure \ref{fig:allparameters_vs_mass}, but for the TNG100 subsamples, where ``TNG100'' in the legend in contrast to ``MO'' used for mock observation in Figure \ref{fig:allparameters_vs_mass}.}  
		\label{fig:allparameters_vs_mass_ori}
	\end{figure*}

	In this section, we explore how the choice and configuration of convolutional kernels in mock observations affect the derived morphological parameters of galaxies. Figure \ref{fig:allparameters_vs_mass_ori} presents the variation of morphological parameters with stellar mass for different types of galaxies in the TNG100 dataset. To facilitate a better comparison with the mock observation galaxies, all classifications of ETGs and LTGs in this section are based on CNN Model 1. It is evident that the morphological parameters of TNG100 galaxies significantly differ from those of SDSS galaxies. The largest discrepancy is observed in the asymmetry parameter, $A$, where TNG100 galaxies are at least 0.2 dex higher than those in SDSS. $R_e$ shows the highest consistency. Furthermore, TNG100 ETGs exhibit significant morphological differences compared to their SDSS counterparts. For the concentration index \( C \), values are 0.5 dex lower than SDSS at the low-mass end and 0.5 dex higher at the high-mass end. For the \( n \), ETGs in TNG100 are 1 dex lower than SDSS at lower masses but closely match the SDSS values at higher masses. Comparing with Figure \ref{fig:allparameters_vs_mass}, it is evident that the consistency of morphological parameters between TNG100 and SDSS is significantly lower than that between mock observations and SDSS. This discrepancy is primarily due to differences in observational conditions, which degrade the data quality by lowering resolution and increasing noise. Reduced resolution causes the central regions of galaxies to be affected by the PSF, leading to a significant decrease in concentration parameters. Noise, on the other hand, obscures the fainter outer regions of galaxies, reducing the overall size and making the image data appear smoother.
	
\begin{figure*}[t]
		\centering
		\includegraphics[width=\textwidth]{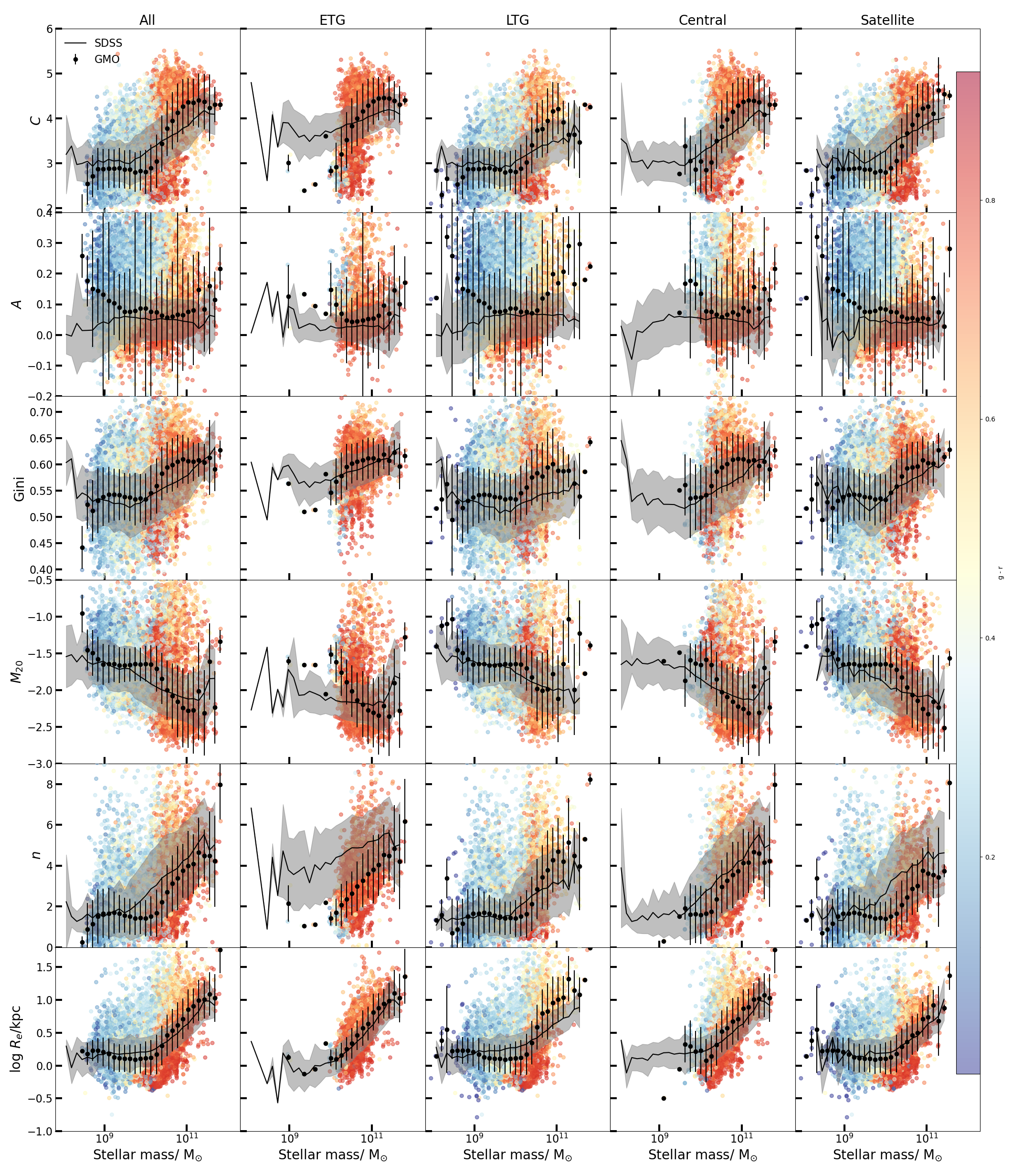}
		\caption{The same as Figure \ref{fig:allparameters_vs_mass}, but for the Gaussian mock observation subsamples, where ``GMO'' in the legend represents Gaussian mock observation, in contrast to ``MO'' used for mock observation in Figure \ref{fig:allparameters_vs_mass}.}  
		\label{fig:allparameters_vs_mass_gaussian}
	\end{figure*}

	In section \ref{Sec_method_mock}, we detailed the data processing steps for mock observations, where the PSF estimation is derived from actual SDSS data. To further investigate the impact of PSF on the estimation of galaxy morphology, we used a Gaussian kernel with FWHM = 0.396$^{\prime\prime}$ for PSF convolution. The resulting sample is referred to as the Gaussian mock observation data. Figure \ref{fig:allparameters_vs_mass_gaussian} presents the variation of morphological parameters with stellar mass for different types of galaxies in the Gaussian mock observation subsample. To ensure a consistent comparison with the mock observation galaxies, the classifications of ETGs and LTGs in this figure are also based on CNN Model 1. Compared to Figure \ref{fig:allparameters_vs_mass_ori}, Figure \ref{fig:allparameters_vs_mass_gaussian} shows that the Gaussian mock observation data exhibit a higher consistency with SDSS results, though still not as high as the consistency shown by the original mock observation in Figure \ref{fig:allparameters_vs_mass}. The Gaussian mock observation data still display a systematic offset of about 0.1 dex in $A$ compared to SDSS galaxies in high and low mass end. $R_e$ shows the highest consistency, with almost all Gaussian mock observation galaxies having sizes consistent with their SDSS counterparts. Consistent with Figure \ref{fig:allparameters_vs_mass_ori}, \( n \) in Gaussian mock observations shows a systematic offset only for ETGs, which are approximately 0.5 dex lower than their SDSS counterparts. This discrepancy increases as stellar mass decreases. Other subsamples exhibit relatively higher consistency with the SDSS \( n \) values.
	
	Comparing results in Figures \ref{fig:allparameters_vs_mass}, \ref{fig:allparameters_vs_mass_ori}, and \ref{fig:allparameters_vs_mass_gaussian}, it is found that, given a sufficiently accurate PSF, the \sersic{} function is a more robust method for estimating galaxy morphology compared to non-parametric methods, as demonstrated in several studies (e.g., \citealt{Boris2013,Davari2014,Gong2023}). Non-parametric statistics, although crucial for galaxy classification, are sensitive to data quality, such as resolution and signal-to-noise ratio (S/N). We observe that $A$ is most affected, with only the mock observation galaxies showing high consistency with SDSS. $C$ is also affected, with its impact depending on the galaxy type; higher $C$ galaxies are more sensitive to data quality. Despite differences in data quality, the distinction between ETGs and LTGs is consistently recovered, with ETGs always occupying high concentration regions and LTGs low concentration regions. Our results emphasize the importance of considering data quality in non-parametric statistics and emphasize that while numerical values may vary, the parameter differences between ETGs and LTGs remain valid. 
	
	\subsection{ Lack of ETG at the Small-Mass End of Dwarf Galaxies}\label{sec_lack_of_etg}
	
	Figure \ref{fig:allparameters_vs_mass} illustrates the comparison between mock observation galaxies and SDSS galaxies across different stellar mass bins for both the $C$ coefficient and $n$. Notably, the figure reveals a pronounced absence of ETGs in the dwarf galaxy regime ($M_* < 10^{10} M_{\odot}$) within mock observations. Within this mass range, all mock observation data points, within their error regions, exhibit characteristics of low concentration ($C \approx 2.67$, $n \approx 1.64$), suggesting that LTGs predominantly occupy the dwarf galaxy regime. The second column of Figure \ref{fig:allparameters_vs_mass} indicates that mock observation ETGs are scarcely represented, constituting only 0.05\% of galaxies in this mass range, with these few data points displaying relatively lower concentrations ($C \approx 2.54$, $n \approx 1.33$). Although labeled as ETGs, these galaxies exhibit LTG-like properties, implying a near absence of ETGs at the dwarf galaxy end in mock observations. In contrast, SDSS data within the dwarf galaxy mass range still contain 19.79\% of ETGs, which demonstrate significantly higher concentrations ($C \approx 3.67$, $n \approx 3.86$), affirming their classification as true ETGs. 
	
	The absence of dwarf ETGs in mock observations could be due to limitations in resolution. \citet{zeng2024} found that most low-mass ($M_* < 10^{8.5-9.0} M_{\odot}$) central galaxies in TNG50 exhibit dispersion-dominated kinematics, aligning well with observational data. Moreover, they emphasized that the impact of resolution must be carefully accounted for when studying the morphology of simulated galaxies, as lower resolutions may hinder the accurate identification of key morphological features, particularly in dwarf galaxies.
	
	Apart from resolution limitations, variations in the formation and evolution mechanisms of simulated galaxies compared to observed galaxies may also lead to morphological differences. Dwarf galaxies, important tracers of environmental build-up history, are particularly susceptible to external factors like tidal harassment (\citealt{Moore1996,Smith2015}) and internal accretion processes (\citealt{Graham2017,Janz2017}). These galaxies can undergo significant mass stripping and morphological transformations. \citet{Donnari2019} found that while the TNG simulations predict a strong bimodality in color distributions, the SFR distributions are unimodal with a peak at the star-forming main sequence (MS). Similarly, \citet{zhao2020} noted that the bimodal specific star formation rate function (sSFRF) seen in SDSS observations is absent in TNG simulations, where only a peak for high star-forming objects is present. \citet{Katsianis2021} further analyzed this issue and suggested that the unimodal SFR distribution in TNG simulations is not solely due to resolution limits. \citet{Corcho2021} emphasized that the significant mismatch in the star formation activity of passive systems between simulations and observations is not due to selection effects, but rather indicates a need for weaker quenching in simulations to maintain low star formation activity. Our findings show the importance of exercising caution when using TNG100 data for morphological studies, as the absence of dwarf ETGs in the simulations could lead to biases in the analysis and conclusions.
	
	\subsection{Comparison with Other Work}
	
	Recent studies have leveraged TNG data to produce mock observations for comparison with actual observations, enriching our understanding of galaxy morphology across simulations and real-observation datasets. \customcite{Rodriguez-Gomez_2019}{R19} generated synthetic images for approximately 27,000 galaxies from both the TNG and the original Illustris simulations, aimed at emulating Pan-STARRS observations for galaxies with stellar masses in the range of $ \log_{10}(M_*/M_{\odot}) \approx 9.8-11.3 $ at $ z \approx 0.05 $. Morphological diagnostics, including Gini-$M_{20}$ and CAS statistics alongside 2D \sersic{} fits. The study noted that optical morphologies of TNG galaxies closely align with observational data, marking an improvement from the original Illustris outputs. Despite this progress, challenges remain in reproducing a strong morphology-colour relation and the morphology-size relation observed in real galaxies. Building on these findings, \customcite{Huertas-Company_2019}{H19} delved deeper into the optical morphologies of TNG galaxies at $ z \approx 0 $, employing a CNN trained on SDSS visual morphologies. By generating mock SDSS images of around 12,000 galaxies from the simulation and adjusting for SDSS $ r $-band characteristics, they processed these through a neural network to classify galaxies into four morphological types. Their analysis corroborated the diversity of optical morphologies within the TNG suite, though discrepancies were noted in the correlation between optical morphology and \sersic{} index, as well as in the stellar mass functions segmented by morphology, especially among late-type galaxies at higher masses.
	
	\begin{figure*}[th]
		\centering
		\includegraphics[width=\textwidth]{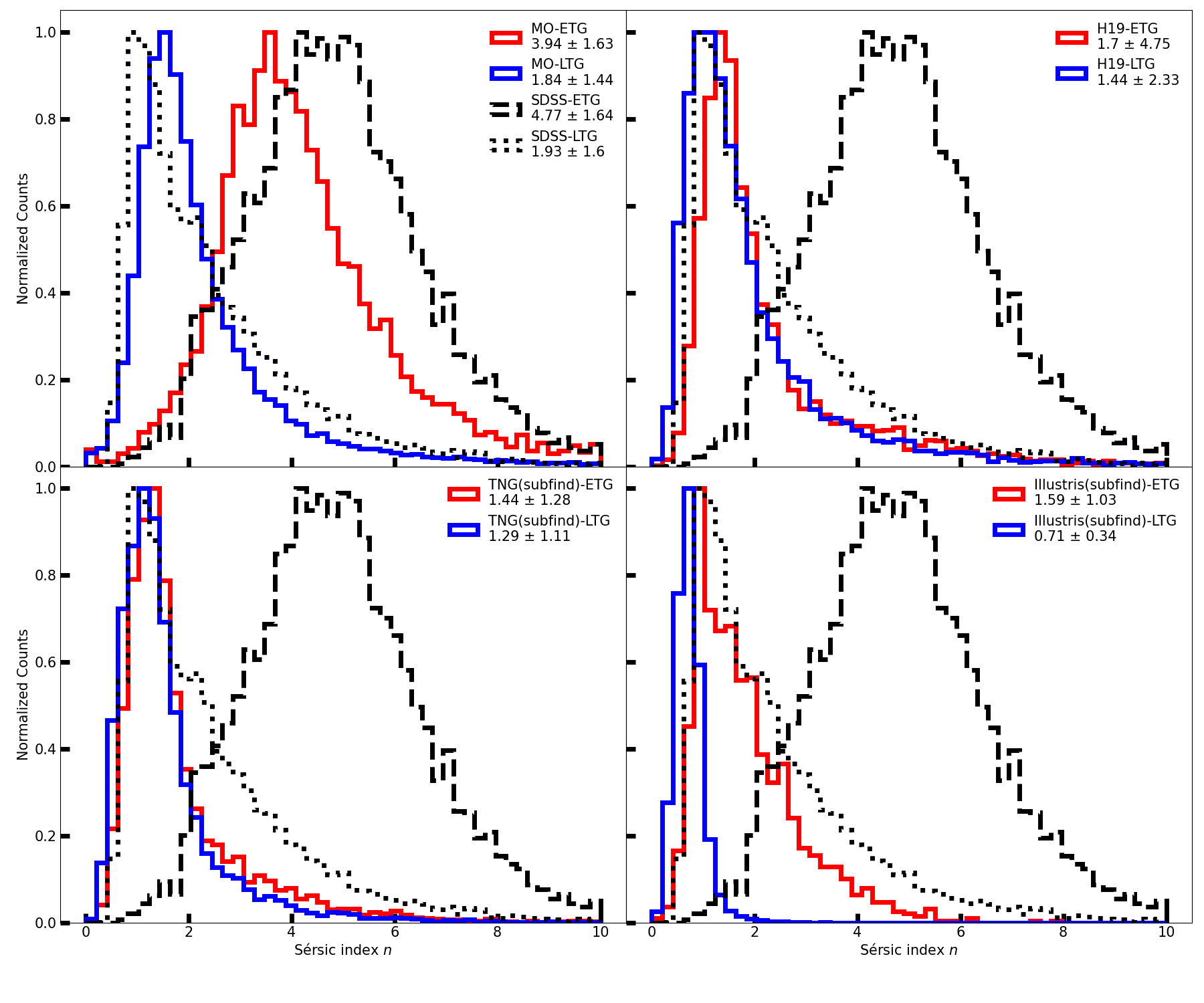}
		\caption{Histograms of the \sersic{} index derived from various datasets, in comparison to those from SDSS. The top-left panel showcases mock observation result from this study, and the results from \customcite{Huertas-Company_2019}{H19} in the top-right panel. The bottom panel elucidates findings from mock observations applied to TNG (bottom-left) and  Illustris (bottom-right) synthetic image data according to \customcite{Rodriguez-Gomez_2019}{R19}, adopting the methodology delineated in our research. Across all panels, ETGs are represented in red and LTGs in blue. Black dashed and dotted lines correspond to SDSS ETGs and LTGs, respectively. Median values along with their standard deviations, indicating error, are annotated in the legend.}
		\label{fig:n_comparison}
	\end{figure*}
	
	\begin{figure*}[th]
		\centering    \includegraphics[width=\textwidth]{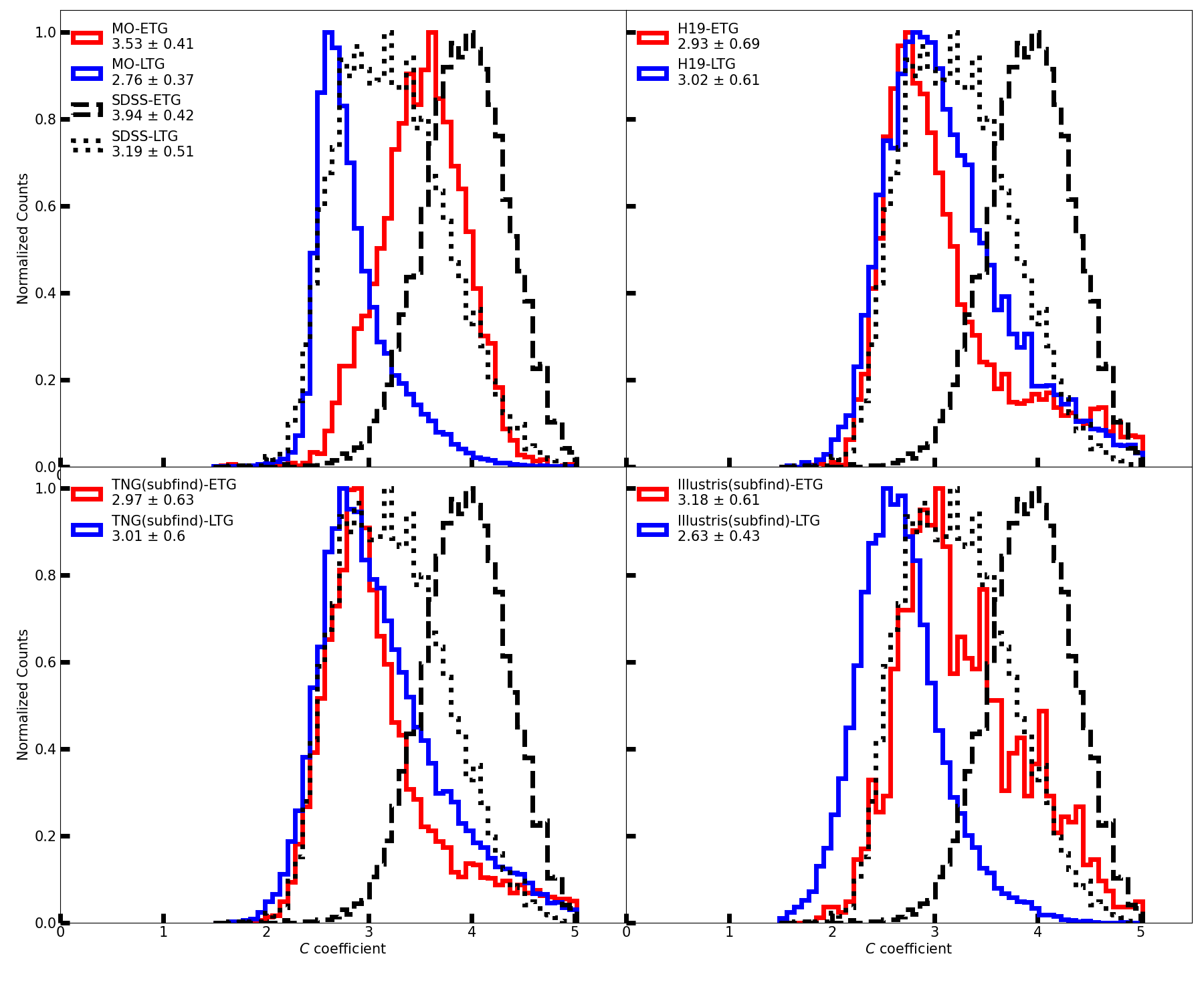}
		\caption{Histograms of the $C$ coefficient derived from various datasets, in comparison to those from SDSS. The format is the same as Figure \ref{fig:n_comparison}.}
		\label{fig:c_comparison}
	\end{figure*}
	
	Given the variation in galaxy redshifts and stellar mass ranges considered in different studies, for a coherent comparison with other works, this section focuses on mock observation and SDSS galaxies with stellar masses exceeding $10^{9.5} M_{\odot}$. The top row panels in Figures \ref{fig:n_comparison} and \ref{fig:c_comparison} contrast $n$ and $C$ coefficient results from this study (upper left) with those from \customcite{Huertas-Company_2019}{H19} (upper right), respectively. Combining observations from both figures, the galaxies in \customcite{Huertas-Company_2019}{H19}, regardless of being ETGs or LTGs, exhibit no significant delineation in concentration parameters ($n \approx 1.6$, $C \approx 3.0$). However, a clear distinction with ETGs showing significantly higher concentrations than LTGs is manifest in our mock observation results. \customcite{Huertas-Company_2019}{H19} suggest that the lack of a pronounced difference between ETGs and LTGs might be attributable to the disparate methodologies applied in calculating the $n$ between simulations and observations.
	
	To delve into the reasons behind the indistinct concentration differences between ETGs and LTGs observed in \customcite{Huertas-Company_2019}{H19}, we adopted the synthetic images from \customcite{Rodriguez-Gomez_2019}{R19}, based on TNG and Illustris simulations, and applied the same mock observation process used in this study to estimate the morphological parameters of galaxies. The primary distinction between these Synthetic Images and the mock observations conducted in our research lies in the definition of galaxies—our approach uses the SBLSP method, while the Synthetic Images utilize the \texttt{subfind} algorithm for galaxy sample identification. The bottom-left panel of Figures \ref{fig:n_comparison} and \ref{fig:c_comparison} illustrates the results based on Synthetic Images for TNG, and the bottom-right panel presents those based on Illustris. Despite employing an identical mock observation process, the results significantly diverge. For TNG samples defined using \texttt{subfind}, the \sersic{} index of ETGs ($n \approx 1.44 \pm 1.28$, $C \approx 2.97 \pm 0.63$) is only marginally higher than that of LTGs ($n \approx 1.29 \pm 1.11$, $C \approx 3.01 \pm 0.60$), a difference too slight to serve as a robust criterion for classifying ETGs and LTGs since this level of concentration falls within the error range of LTGs' concentration observed in SDSS. This result emphasizes that compared to the mock observation techniques or the methods employed for calculating morphological parameters, consistency in the definition of galaxy samples is more important for replicating morphological parameters in observation accurately. This is attributed to the conventional data processing practice where galaxy sample delineation typically relies on luminosity distinctions, as exemplified by segmentation software like SExtractor (\citealt{Bertin1996}).
	
	Conversely, galaxies subjected to mock observations based on the Illustris simulation exhibited relatively superior concentration results. Notably, ETGs ($n \approx 1.59 \pm 1.03$, $C \approx 3.18 \pm 0.61$) demonstrated significantly higher concentrations than LTGs ($n \approx 0.71 \pm 0.34$, $C \approx 2.63 \pm 0.43$), with both parameters revealing a robust bimodal distribution. No matter what method is used to estimate galaxy morphology, it was observed that galaxies from numerical simulations consistently display lower concentrations compared to those in the SDSS. This discrepancy is particularly pronounced among ETGs, where SDSS ETGs predominantly cluster around $n \approx 5$ or $C \approx 4$, whereas the principal distribution of ETGs in all numerical simulations tends towards $n \approx 3$ or $C \approx 3.6$. Several factors might contribute to this phenomenon, including the insufficient resolution affecting the stellar particles in numerical simulations or inadequate representation of feedback mechanisms. This discrepancy stresses the challenges in replicating the intricate details of galaxy morphology through simulations, emphasizing the necessity for ongoing refinement of simulation techniques to bridge the gap with observational data.
	
	\section{Summary}\label{Sec_summary}
	
	This work presents a comprehensive morphological analysis comparing mock observations generated from the TNG100 simulation with actual observations from the SDSS. We generated mock observations for approximately 34,000 galaxies with stellar masses exceeding $10^8 M_{\odot}$ across various redshifts (z=0.01, 0.1, 0.2) and viewing planes ($x\text{-}y, y\text{-}z, z\text{-}x$) by applying comprehensive data processing techniques including PSF convolution, background addition, and noise simulation to the TNG100 simulation data. Crucially, we employed the SBLSP algorithm for brightness-based galaxy sample definition, designed to mimic observational segmentation methods. Subsequently, we implemented a hierarchical deep learning approach using three CNNs to classify these mock-observed galaxies into four distinct morphological classes based on \textnormal{T-type}: ellipticals ($\textnormal{T-type} \leq -3$), S0/a ($-3 < \textnormal{T-type} \leq 0$), Sab/Sb ($1 \leq \textnormal{T-type} < 4$), and Sc/Sd/irregulars ($\textnormal{T-type} \geq 4$), achieving high classification accuracy.
	
	Our analysis shows that the TNG100 mock observations successfully reproduce key morphological features seen in the SDSS dataset. Notably, our results capture the distinct parameter distributions distinguishing ETG and LTG, often exhibiting bimodality, and correctly recover the systematic variation of parameters like concentration across the galaxy morphological type. However, despite this strong general agreement, we identified two specific areas of divergence. Firstly, there is a significant lack of ETGs in the dwarf galaxy regime ($M_* < 10^{10}M_{\odot}$) within the mock observations (0.05\% vs. 19.79\% in SDSS). Secondly, the morphological parameters derived for Sab/Sb and Sc/Sd/irregular galaxies in the mocks show minimal distinction, contrary to the clear differences observed in real galaxies. These discrepancies might stem partly from inherent properties of the TNG100 simulation, such as its potentially elevated star formation efficiency and its resolution limits. We are currently conducting similar analyses using the higher-resolution TNG50 simulation to further investigate the origins of these differences. Nevertheless, the overall success in matching broad morphological trends underscores the value of our approach.
	
	Compared to prior studies like \customcite{Huertas-Company_2019}{H19}, our approach yields a marked improvement in distinguishing galaxy types. The analysis by \customcite{Huertas-Company_2019}{H19}, likely influenced by their use of gravity-based galaxy sample definition, resulted in similar morphological parameter distributions for both ETGs and LTGs, failing to capture the clear distinctions seen in observations. We demonstrate that the improved recovery in our work is primarily attributable to the adoption of the SBLSP brightness-based sample definition, contrasting sharply with the results obtained using those gravity-based definitions. Our findings thus highlight that achieving consistency in how galaxy samples are defined between simulations and observations is as critical as applying realistic mock imaging for robust comparisons. Separately, we also assessed the impact of observational effects (PSF, noise, background) on the classification process itself. We found that these factors alone could potentially cause our CNN model to misclassify the morphology of approximately 7.87\% of the galaxies. The full catalog is available in the online article.
	
	\section*{ACKNOWLEDGMENTS}\label{ACKNOWLEDGMENTS}
	
	We thank the anonymous referee for helpful suggestions and constructive comments to improve the manuscript. The authors express their gratitude to the Illustris and TNG projects for providing the simulation data. This work was supported by the National Natural Science Foundation of China (Nos. 12073089, 12003079) and the China Manned Space Program through its Space Application System. L.T. is also supported by Natural Science Foundation of Sichuan Province (No. 2022NSFSC1842), the Fundamental Research Funds of China West Normal University (CWNU, No. 21E029), and the Sichuan Youth Science and Technology Innovation Research Team (21CXTD0038). Most computations were performed on the Kunlun HPC at the School of Physics and Astronomy, Sun Yat-sen University. This research has made use of data from SDSS-III, which is funded by the Alfred P. Sloan Foundation, the Participating Institutions, the National Science Foundation, and the U.S. Department of Energy Office of Science; the SDSS-III web site is http://www.sdss3.org/. This research has also made use of the SAO/NASA’s Astrophysics Data System Bibliographic Services. This work is supported by China Manned Space Program through its Space Application System.
	
	\appendix 
	\begin{table*}[t]
	\centering
	\caption{Parameters catalog for mock observation galaxy}
	\label{tab:catalog}
	\begin{tabular}{@{}cccccccccc@{}}
		\toprule
		Galaxy\_ID & Gini & $M_{20}$ & $C$ & $A$ & $S$ & Type & x\_sersic & y\_sersic & $R_{e}$\_sersic \\
		$n$\_sersic & q\_sersic & PA\_sersic & RFF & EVI & x\_iso & y\_iso & sma\_iso & e\_iso & PA\_iso \\
		g\_absmag & r\_absmag & i\_absmag & z\_absmag & CNN\_N10 & LTG\_Pro\_N10 & CNN\_TNG & LTG\_Pro\_TNG & $r_{20}$ & $r_{50}$ \\
		$r_{80}$ & $r_{90}$ & groupid & plane & redshift & Mass\_aper & SFR\_aper & Age\_aper & Met\_aper & E/S0 \\ 
		P\_S0 & Sab/Scd & P\_Scd \\
		\midrule
		2123 & 0.49 & -1.73 & 2.67 & 0.09 & 0.07 & 1 & 1547.50 & 902.60 & 3.67 \\
		1.41 & 0.76 & 64.74 & 0.00 & 0.08 & 1546.87 & 901.51 & 11.96 & 0.13 & 62.17 \\
		-21.29 & -21.59 & -21.78 & -21.92 & 1 & 0.84 & 1 & 1.00 & 2.35 & 4.73 \\
		8.34 & 11.11 & 0 & xy & 0.01 & 0.79 & 2.09 & 1.93 & 0.02 & 1 \\ 
		0.71 & 0 & 0.20 \\
		\midrule
		7317 & 0.59 & -2.17 & 3.73 & 0.01 & 0.02 & 1 & 2133.03 & 2026.49 & 11.68 \\
		4.89 & 0.36 & 84.12 & -0.04 & -0.04 & 2131.67 & 2025.35 & 30.43 & 0.59 & -89.80 \\
		-21.71 & -22.45 & -22.79 & -23.05 & 0 & 0.29 & 0 & 0.10 & 4.03 & 10.10 \\
		21.75 & 29.27 & 0 & xy & 0.01 & 5.14 & 0.02 & 5.80 & 0.02 & 1 \\ 
		0.93 & 0 & 0.17 \\
		\midrule
		33372 & 0.52 & -1.76 & 2.78 & 0.01 & 0.00 & 1 & 3124.78 & 2980.45 & 2.10 \\
		2.24 & 0.75 & 66.34 & -0.05 & -0.14 & 3123.92 & 2979.34 & 9.61 & 0.16 & 36.31 \\
		-20.27 & -20.84 & -21.15 & -21.38 & 1 & 0.87 & 1 & 1.00 & 1.75 & 3.50 \\
		6.97 & 9.63 & 2 & xy & 0.01 & 1.01 & 0.34 & 5.19 & 0.02 & 1 \\
		0.69 & 0 & 0.34 \\
		\midrule
		40453 & 0.47 & -1.67 & 2.65 & -0.01 & 0.00 & 1 & 2518.07 & 1967.08 & 2.25 \\
		1.80 & 0.54 & -57.65 & -0.11 & -0.15 & 2516.92 & 1966.01 & 6.44 & 0.20 & -51.30 \\
		-18.93 & -19.37 & -19.64 & -19.84 & 1 & 0.98 & 1 & 1.00 & 1.77 & 3.49 \\
		6.41 & 8.46 & 3 & xy & 0.01 & 0.25 & 0.16 & 5.22 & 0.02 & 1 \\ 
		0.72 & 1 & 0.67 \\
		\midrule
		65388 & 0.54 & -1.88 & 3.23 & 0.17 & 0.05 & 1 & 5270.74 & 2008.12 & 4.15 \\
		4.58 & 0.47 & -58.50 & 0.01 & 0.21 & 5269.37 & 2008.81 & 13.78 & 0.27 & -62.20 \\
		-21.25 & -21.42 & -21.53 & -21.67 & 1 & 0.86 & 1 & 1.00 & 2.00 & 4.41 \\
		9.15 & 12.35 & 7 & xy & 0.01 & 0.58 & 3.12 & 2.13 & 0.01 & 1 \\ 
		0.73 & 0 & 0.26 \\
		\midrule
		65632 & 0.60 & -2.32 & 3.99 & 0.00 & 0.00 & 1 & 5357.57 & 1708.85 & 17.30 \\
		4.70 & 0.70 & -77.10 & -0.02 & 0.28 & 5355.92 & 1708.01 & 50.79 & 0.43 & -79.82 \\
		-23.22 & -24.01 & -24.40 & -24.68 & 0 & 0.00 & 0 & 0.00 & 5.01 & 14.67 \\
		34.41 & 46.86 & 7 & xy & 0.01 & 38.17 & 1.95 & 10.94 & 0.02 & 0 \\
		0.04 & 0 & 0.11 \\
		\midrule
		107658 & 0.52 & -1.66 & 2.71 & 0.02 & 0.00 & 1 & 4029.21 & 1024.91 & 2.73 \\
		1.51 & 0.61 & -71.89 & -0.04 & -0.07 & 4028.53 & 1024.00 & 10.29 & 0.30 & -67.12 \\
		-20.05 & -20.48 & -20.72 & -20.95 & 1 & 0.90 & 1 & 1.00 & 1.99 & 3.93 \\
		8.04 & 11.13 & 16 & xy & 0.01 & 0.59 & 0.51 & 5.13 & 0.02 & 1 \\ 
		0.72 & 1 & 0.52 \\
		\midrule
		115403 & 0.51 & -1.70 & 2.85 & 0.02 & 0.00 & 1 & 2099.39 & 2892.81 & 2.97 \\
		2.58 & 0.43 & -0.32 & -0.04 & -0.11 & 2098.36 & 2891.46 & 11.45 & 0.38 & 0.97 \\
		-20.61 & -21.02 & -21.23 & -21.39 & 1 & 0.87 & 1 & 1.00 & 2.00 & 4.10 \\
		7.86 & 11.04 & 19 & xy & 0.01 & 0.55 & 1.15 & 2.32 & 0.02 & 1 \\ 
		0.75 & 0 & 0.28 \\
		\midrule
		215058 & 0.58 & -2.10 & 3.73 & -0.01 & 0.01 & 1 & 1688.37 & 1098.79 & 8.69 \\
		5.01 & 0.74 & 13.66 & -0.07 & -0.11 & 1687.24 & 1098.74 & 22.66 & 0.39 & 19.20 \\
		-21.25 & -22.00 & -22.34 & -22.60 & 0 & 0.43 & 0 & 0.13 & 3.07 & 8.24 \\
		17.72 & 23.92 & 109 & xy & 0.01 & 4.27 & 0.01 & 8.57 & 0.02 & 1 \\
		0.96 & 0 & 0.21 \\
		\bottomrule
	\end{tabular}
	\footnotesize{}
\end{table*}
	In this appendix, we provide a comprehensive table detailing a variety of parameters derived from our analysis. The full catalog is available in the online article. Below is a brief description of each parameter included in the table:
	
	\begin{itemize}
		\item \textbf{Galaxy\_ID}: A unique identifier for each galaxy in the dataset. It is important to note that this ID does not directly correspond to the subfind\_ID provided on the TNG official website, due to additional processing and selection criteria applied in this study.
		\item \textbf{Gini}: The Gini coefficient.
		\item \textbf{$M_{20}$}: The second-order moment of the brightest 20\% of the galaxy's light.
		\item \textbf{$C$}: The concentration index.
		\item \textbf{$A$}: The asymmetry index.
		\item \textbf{$S$}: The smoothness (or clumpiness) index.
		\item \textbf{Type}: Galaxy type, where 0 denotes a central galaxy and 1 signifies a satellite galaxy.
		\item Parameters \textbf{x\_sersic}, \textbf{y\_sersic}, \textbf{n\_sersic}, \textbf{$R_{e}$\_sersic},
		\textbf{q\_sersic} and \textbf{PA\_sersic} represent the centroid coordinates in x and y, \sersic{} index, effective radius, axis ratio, and position angle in group image derived from \sersic{} fitting, respectively.
		\item \textbf{RFF}: The residual flux fraction, a measure of the fraction of light not accounted for by the \sersic{} model.
		\item \textbf{EVI}: The environmental density index, assessing the local density of galaxies around the target galaxy. \textbf{RFF} and \textbf{EVI}, which are defined in \cite{Hoyos2011}, are used to determine whether the \sersic{} model is an adequate fit.
		\item Parameters \textbf{x\_iso}, \textbf{y\_iso}, \textbf{sma\_iso}, \textbf{e\_iso}, and \textbf{PA\_iso} represent the centroid coordinates in x and y, semi-major axis length, ellipticity, and position angle in group image derived from isophotal fitting, respectively.
		\item Absolute magnitudes form aperture photometry in $g$, $r$, $i$, and $z$ bands are denoted as \textbf{g\_absmag}, \textbf{r\_absmag}, \textbf{i\_absmag}, and \textbf{z\_absmag}.
		\item \textbf{CNN\_N10}, \textbf{CNN\_TNG}, \textbf{E/S0}, and \textbf{Sab/Scd} indicate the classification results from CNN models trained on different datasets. \textbf{CNN\_N10} and \textbf{CNN\_TNG} correspond to Model 1 trained on the \customcite{Nair_2010}{N10} and TNG100 datasets, respectively, with 0 for ETGs and 1 for LTGs. \textbf{E/S0} represents Model 2's classification, where 0 denotes an elliptical galaxy (E) and 1 denotes an S0 galaxy. Similarly, \textbf{Sab/Scd} represents Model 3's classification, where 0 corresponds to an Sab/Sb galaxy and 1 to an Scd/Sd/irregular galaxy.  
		\item \textbf{LTG\_Pro\_N10}, \textbf{LTG\_Pro\_TNG}, \textbf{P\_S0}, and \textbf{P\_Scd} represent the classification probabilities from the respective CNN models. \textbf{LTG\_Pro\_N10} and \textbf{LTG\_Pro\_TNG} indicate the probability of a galaxy being classified as an LTG by Model 1 trained on the N10 and TNG100 datasets, respectively. \textbf{P\_S0} denotes the probability of a galaxy being classified as an S0 galaxy by Model 2, with higher values suggesting a stronger likelihood of this classification. Similarly, \textbf{P\_Scd} represents the probability of a galaxy being classified as an Scd/Sd/irregular galaxy by Model 3.  
		\item \textbf{$r_{20}$}, \textbf{$r_{50}$}, \textbf{$r_{80}$}, \textbf{$r_{90}$}: These parameters represent radii enclosing 20\%, 50\% (also known as the effective radius), 80\%, and 90\% of a galaxy's total light, respectively. These radii are critical for understanding the light distribution and concentration of galaxies.
		\item \textbf{groupid}: This parameter identifies the galaxy group to which a galaxy belongs, as determined by FoF algorithm.
		\item \textbf{plane}: Projection plane. This can be, for example, the $x\text{-}y, y\text{-}z, z\text{-}x$ plane of the simulation volume, affecting the apparent morphology of galaxies.
		\item \textbf{redshift}: The redshift value of the galaxy.
		\item Parameters with the subscript \textbf{\_aper}, such as \textbf{Mass\_aper}, \textbf{SFR\_aper}, \textbf{Age\_aper}, and \textbf{Met\_aper}, correspond to stellar mass, star formation rate, age, and metallicity measured within an aperture defined by a surface brightness of 26.5 mag arcsec$^{-2}$.
	\end{itemize}
	
	\clearpage
	
	\bibliography{revised_export-bibtex}{}
	\bibliographystyle{aasjournal}
	
\end{document}